\documentclass[prd,nofootinbib,superscriptaddress]{revtex4}

\usepackage{amsfonts,amssymb,amsthm,bbm,graphicx,mathtools}
\usepackage{subcaption}
\usepackage{amsmath}

\DeclareMathOperator\arcsinh{arcsinh}
\usepackage{array}
\usepackage{hyperref}

\usepackage{amsbsy}

\newcommand*\Bell{\ensuremath{\boldsymbol\ell}}

\usepackage{subcaption}
\usepackage{color}
\usepackage{tikz}
\usepackage[compat=1.1.0]{tikz-feynman}
\usetikzlibrary{arrows}
\usetikzlibrary{calc}
\usetikzlibrary{decorations.pathmorphing}
\usetikzlibrary{decorations.markings}
\usepackage{bigints}

\newcommand{\N}{{\mathbb N}}
\newcommand{\R}{{\mathbb R}}
\newcommand{\Z}{{\mathbb Z}}

\newcommand{\cC}{{\mathcal C}}
\newcommand{\cD}{{\mathcal D}}

\newcommand{\cG}{{\mathcal G}}
\newcommand{\cH}{{\mathcal H}}

\newcommand{\cK}{{\mathcal K}}
\newcommand{\cL}{{\mathcal L}}

\newcommand{\cR}{{\mathcal R}}
\newcommand{\cS}{{\mathcal S}}

\newcommand{\cU}{{\mathcal U}}

\newcommand{\SU}{\mathrm{SU}}

\newcommand{\SL}{\mathrm{SL}}
\newcommand{\SO}{\mathrm{SO}}

\newcommand{\rL}{\mathrm{L}}
\newcommand{\rE}{\mathrm{E}}

\newcommand{\su}{{\mathfrak{su}}}

\newcommand{\hu}{{\hat{u}}}

\def\be#1\ee{\begin{equation}#1\end{equation}}
\def\beq#1\eeq{\begin{eqnarray}#1\end{eqnarray}}
\def\bea#1\eea{\begin{align}#1\end{align}}
\def\nn{\nonumber}

\newcommand{\f}{\frac}

\newcommand{\ra}{\rangle}

\newcommand{\tr}{{\mathrm{Tr}}}

\def\rd{\textrm{d}}

\newcommand{\id}{\mathbb{I}}

\def\vphi{\varphi}
\def\eps{\epsilon}
\def\ka{\kappa}
\def\om{\omega}

\def\vP{\vec{P}}
\def\vx{\vec{x}}
\def\kEphi{ {\mathbf k}^{E}_\varphi}

\def\vJ{\vec{J}}
\def\ha{\hat{a}}
\def\tdelta{\widetilde{\delta}}
\def\tmu{\tilde{\mu}}
\newcommand{\bk}{{\mathbf{k}}}
\def\deltaH{ \delta^{\mathrm{H}}}
\def\sh{\mathrm{h}}
\def\sk{\mathrm{k}}

\def\centerarc[#1](#2)(#3:#4:#5)% Syntax: [draw options] (center) (initial angle:final angle:radius)
{ \draw[#1] ($(#2)+({#5*cos(#3)},{#5*sin(#3)})$) arc (#3:#4:#5); }

\def\centerarcnodes[#1](#2)(#3:#4:#5)(#6,#7)% Syntax: [draw options] (center) (initial angle:final angle:radius)
{\coordinate(#6) at ($(#2)+({#5*cos(#3)},{#5*sin(#3)})$);
	\coordinate(#7) at ($(#2)+({#5*cos(#4)},{#5*sin(#4)})$);
	\draw[#1] ($(#2)+({#5*cos(#3)},{#5*sin(#3)})$) arc (#3:#4:#5); }

\def\angcircle(#1)(#2)(#3:#4) {\coordinate(#1) at ($(#2)+({#4*cos(#3)},{#4*sin(#3)})$); }

\tikzset{->-/.style={decoration={
			markings,
			mark=at position #1 with {\arrow{>}}},postaction={decorate}}}

\tikzset{-<-/.style={decoration={
			markings,
			mark=at position #1 with {\arrow{<}}},postaction={decorate}}}
   
\newcolumntype{M}[1]{>{\raggedright}m{#1}}

\usepackage{tikz-3dplot}
\usetikzlibrary{3d}
 \usetikzlibrary{arrows, decorations.markings, calc, fadings, decorations.pathreplacing, patterns, decorations.pathmorphing, positioning}

\usepackage{pgfplots}
\pgfplotsset{compat=1.18}
\begin{document}

\title{Matter coupled to 3d Quantum Gravity: One-loop  Unitarity}

\author{{\bf Etera R. Livine}}
\email{etera.livine@ens-lyon.fr}
\affiliation{Univ de Lyon, ENS de Lyon, Laboratoire de Physique, CNRS UMR 5672, Lyon 69007, France}
\author{{\bf Valentine Maris}}
\email{valentine.maris@ens-lyon.fr}
\affiliation{Univ de Lyon, ENS de Lyon, Laboratoire de Physique, CNRS UMR 5672, Lyon 69007, France}
\affiliation{Orsay, IJCLab, Laboratoire de Physique, UMR 9012, Orsay 91400, France}

%\affiliation{Perimeter Institute for Theoretical Physics, 31 Caroline Street North, Waterloo, Ontario, Canada N2L 2Y5}

\date{\today}

\begin{abstract}

We expect quantum field theories for matter to acquire intricate corrections due to their coupling to quantum fluctuations of the gravitational field. This can be precisely worked out in 3d quantum gravity: after integrating out quantum gravity, matter fields are effectively described as noncommutative quantum field theories, with quantum-deformed Lorentz symmetries. An open question remains: Are such theories unitary or not? On the one hand, since these are effective field theories obtained after integrating out high energy degrees of freedom, we may expect the loss of unitarity. On the other hand, as rigorously defined field theories built with Lorentz symmetries and standing on their own, we naturally expect the conservation of unitarity. In an effort to settle this issue, we explicitly check unitarity for a scalar field at one-loop level in both Euclidean and Lorentzian space-time signatures.
We find that unitarity requires adding an extra-term to the propagator of the noncommutative theory, corresponding to a massless mode and given by a representation with vanishing Plancherel measure, thus usually ignored in spinfoam path integrals for quantum gravity. This indicates that the inclusion of matter in spinfoam models, and more generally in quantum gravity, might be more subtle than previously thought.

\end{abstract}

\maketitle
%%%%%%%%%%%%%%%%%%%%%%%%%%%%%%%%%%%%%%%%%%%%%%%%%%%%%%%%
\tableofcontents

%%%%%%%%
\section*{Introduction}
%%%%%%%%

%A comprendre et caser qq part  \textcolor{red}{blabla degree de lib, pas de locaux, mais des globaux (fini), theo topologique.} (se rensigner plus sur les theo topo + deg de lib) Explication, topo = que des degree de lib globaux ie ne depend que de la topo de la variété qui elle n'est pas forcement fixe mais peut changer, c'est eux qu'ils faut regarder  \\

% \cite{Oriti_2001}

Three-dimensional gravity has the crucial property to be an integral system, making it exactly soluble \cite{Witten:1988hc}. More precisely, it is a topological theory, with no local degree of freedom (i.e. no graviton or gravitational waves) but only global freedom degrees coming from the manifold's boundaries and topology. The direct consequence is that the theory can be explicitly and exactly quantized. There are several paths to this 3d quantum gravity, which together form a consistent picture of a topological quantum field theory. Nevertheless, despite being simpler than gravity in four dimensions or higher, 3d quantum gravity is clearly non-trivial and many aspects of the theory still remain to be explored. A specially relevant corner is the coupling of quantum matter to quantum gravity. Being exactly quantizable makes 3d quantum gravity the perfect toy model to gain insight on this key question of quantum gravity.

One path to 3d quantum gravity is the spinfoam quantization of 3d gravity (see \cite{Perez_2003,doná2010introductory,livine2024spinfoam} for introductional reviews of the spinfoam formalism) written as a gauge theory in its first order formulation \`a la Cartan, in terms of vierbein and Lorentz connection. It realizes a direct quantization of the theory as a topological path integral over discrete 3d geometries, leading to the Ponzano-Regge state-sum  \cite{ponzano1968semiclassical}, when the cosmological constant vanishes, and to the Turaev-Viro invariant  \cite{Turaev:1992hq,Turaev:2010pp}, which extends it to a non-vanishing cosmological constant. It provides a regular discrete picture of the quantum space-time at the Planck scale, defining covariant transition amplitudes for the spin network states of 3d loop quantum gravity \cite{Rovelli:1993kc}. It is explicitly related to the Chern-Simons quantization, both at the path integral level \cite{Freidel:2004nb} or at the Hamiltonian level \cite{Noui:2004iy,Bonzom:2014bua,Dupuis:2019yds}, and 't Hooft polygonal quantization \cite{Kadar:2004im}. Moreover its semi-classical regime is understood as quantum Regge calculus \cite{Regge:2000wu}.
%
%At classical and quantum levels, three-dimensional gravity has the property to be an integral system that is exactly soluble \cite{Witten:1988hc}. Furthermore, it is a topological theory, it has no local degree of freedom (i.e., no graviton or gravitational waves) but only global freedom degrees, related to the manifold topology and its borders. There have been several approaches to quantized 3d gravity; one of them is Loop Quantum Gravity (LQG)\cite{rovelli1998strings}. The canonical version of LQG is based on the ADM formulation of gravity, which uses the metric as a variable. Its quantization gives the Wheeler-DeWitt equations. However, the theory was ill-defined, and solutions were found in LQG using others variables, a connection, and a tetrad instead of the metric. Here, we are interested in a covariant version of LQG called the spinfoam approach. It is also based on the connection and tetrad formalism but uses a discretized path integral formulation \cite{Perez_2003} \cite{doná2010introductory}.
%

Of particular interest is the coupling of matter field to spinfoam path integrals. Indeed, it was shown that starting with the coupled system of matter fields and 3d quantum gravity, then integrating over the gravitational degrees of freedom leads to an effective noncommutative quantum field theory (NCQFT) for matter \cite{Freidel:2005me}.The noncommutativity is controlled by the deformation parameter $ \kappa =1/4 \pi G$, identified as the 3d Planck mass (or inverse Planck length) in standard quantum field theory (QFT) conventions with $\hbar=1$ and $c=1$. The noncommutativity effectively deforms the Lorentz group and Poincar\'e symmetry, so that $\kappa^{-1}$ becomes a universal scale invariant under change of observers. Deformation of the symmetries, braiding and noncommutative star-products \cite{Dupuis_2012,Galluccio_2008} are well-understood mathematically. However, several questions still need to be explored and addressed, such as the fate of renormalizability, implementation of Yang-Mills gauge symmetries, causality and unitarity.  
%
%Coupling 3d spinfoam model with matter field and integrating over global gravitational degrees of freedom gives an effective noncommutative quantum field theory for matter \cite{Freidel:2005me}. This effective quantum field theory is characterized by a deformation parameter $\kappa$, which encodes the quantum gravity fluctuation and deforms the usual Poincare spaces. This effective QFT is symmetric under the deformed Poincaré symmetries . It is related to Newton's gravitational constant by $\kappa = \frac{1}{4 \pi G}$ and scale as Planck mass. This deformation parameter is supposed to be a new universal constant of Nature, as the speed of light, and should encode $G$, Newton's constant for gravity, in relevant regime. Several questions still need to be answered, such as the fate of renormalizability, implementation of Yang-Mills gauge symmetry, or unitarity. 
%

Indeed, unitarity is an essential property of QFTs, reflecting the probability flow and the conservation of  information. A consistent quantum field theory should a priori be unitary. A non-unitary QFT does not seem fundamental or physically-relevant. Unfortunately, the investigation of unitarity in noncommutative geometry has been rather limited up to now.
We point out two studies,  \cite{Gomis:2000zz} and \cite{Imai:2000kq,Sasai:2009jm}, performing a one-loop analysis using Cutksoky cutting rules. The first work studies the Moyal space, and shows that unitarity is maintained only if noncommutativity does not affect time. The second line of work studies a scalar field living in the same 3d noncommutative geometry as arising from 3d spinfoam models, that is a three-dimensional noncommutative Lie algebraic spacetime whose coordinate satisfy $\su(2)$ commutator relations, $[x^i,x^j]= {i \kappa^{-1}  \epsilon^{ijk}x_k}$.
%\label{eq:ncrelations}.
Using group momentum space based on the group $\SO(2,1)\sim\SL(2,\R)/\Z_2$, the authors of  \cite{Imai:2000kq} find that the Cutkosky rule is satisfied if and only if the physical mass $M$ is less than $\kappa /\sqrt{2}$ (corresponding to a bare mass $m=\pi \kappa /4$.
%\textcolor{red}{ physical mass $m$ is less than $\pi \kappa /4$}.
%
Here, we wish to push this analysis further and investigate in more details the fate of unitarity for this noncommutative space-time $\mathbb{R}_\kappa^{3}$, especially using the $\SL(2,\R)$ group instead of $\SO(2,1)$, with the goal of establishing (or disproving) the unitarity of effective NCQFTs for matter fields coupled to 3d quantum gravity as derived from the path integral formalism of \cite{Freidel:2005me}. 
%
%The motivation is to study the unitarity of effective noncommutative matter field coupled to three-dimensional quantum gravity introduced in \cite{Freidel:2005me}. 

%\cite{Sasai:2009jm}

\bigskip

More precisely, the Ponzano-Regge state-sum realizes a well-defined path integral  of 3d gravity with vanishing cosmological constant, formulated in terms of a triad 1-form $e$ and a Lorentz connection $\om$. The action is:
\be
S[e,\omega] = \frac{1}{16 \pi G} \int \tr( e\wedge F)
\qquad\textrm{with}\quad
F = \rd \omega + \omega \wedge \omega
\,.
\ee
Without particles, space-time remains flat, $F[w] = 0$, and torsion-free, $d_\omega e = 0$.
Mass and momentum create curvature, while spin is a source of torsion, so that particles arise as topological defects, creating conical singularities \cite{Deser:1983tn,tHooft:1993jwb,tHooft:1996ziz}.
At the quantum level, the original model first developed in 1968 by Ponzano and Regge \cite{ponzano1968semiclassical}, was revisited in the 2000's the context of spinfoams. It was provided with a suitable gauge fixing procedure to make it finite and well-defined \cite{Freidel:2004vi,Barrett:2008wh,Bonzom:2012mb}, leading to the proof of its equivalence with the Ray-Singer analytical torsion and Reidemeister torsion \cite{Barrett:2008wh,Bonzom:2010zh,Bonzom:2011br}. Observables and matter insertion were consistently included as would-be gauge degrees of freedom \cite{Freidel:2004vi,Freidel:2005bb,Barrett:2011qe}.

\begin{figure}[h!]

\centering

\begin{tikzpicture}[scale=1.1]

\coordinate(a) at (0,0) ;
\coordinate(ab) at (1.25,0) ;
\coordinate(b) at (2.5,0);
\coordinate(bc) at (1.75,1.5) ;
\coordinate(ac) at (.5,1.5);
\coordinate(c) at (1,3);
\coordinate(ad) at (1.5,.5);
\coordinate(bd) at (2.75,.5);
\coordinate(cd) at (2,2);
\coordinate(d) at (3,1);

%\coordinate(C) at (2,.33);
%\draw[thick] (C)--(ab);
%\draw[thick] (C)--(bd);
%\draw[thick] (C)--(ad);
%
%\coordinate(A) at (2.3,1.3);
%\draw[thick] (A)--(bc);
%\draw[thick] (A)--(cd);
%\draw[thick] (A)--(bd);
%
%\coordinate(B) at (1.2,1.5);
%\draw[very thick,dotted] (B)--(ac);
%\draw[very thick,dotted] (B)--(ad);
%\draw[very thick,dotted] (B)--(cd);
%
%\coordinate(D) at (1,1);
%\draw[thick] (D)--(ab);
%\draw[thick] (D)--(bc);
%\draw[thick] (D)--(ac);

\draw (a)--(b)--(c)--(a);
\draw[dotted] (a)--(d);
\draw (b)--(d)--(c);

\draw(ab)node[below]{$j_{1}^T$} ;
\draw(ac)node[left]{$j^T_{2}$} ;
\draw(bc)node[ right]{$j^T_{3}$} ;
\draw(ad)++(0.15,.23) node{$j^T_{6}$} ;
\draw(bd)node[right]{$j^T_{5}$} ;
\draw(cd)node[above right]{$j^T_{4}$} ;

\end{tikzpicture}

    \caption{Tetrahedron $T$: the edges $e$ are dressed with $\SU(2)$ representations $j_e^T\in\f\N2$ giving their quantized length in Planck length unit.}
    \label{fig:tetrahedron}

\end{figure}
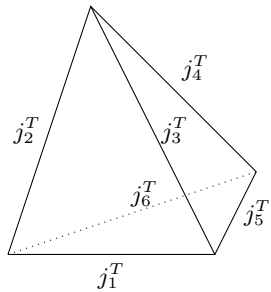

In a nutshell, the Ponzano-Regge model defines 3d quantum gravity amplitudes associated to dressed triangulations. 
In Euclidean signature (+,+,+), considering a finite 3d triangulation $\Delta$ (of arbitrary topology), made of tetrahedra glued together through their triangles, we dress its edges with half-integers $j_{e}\in\f\N2$, which gives their quantized lengths in Planck length unit, $\ell_{e}=j_{e}l_{P}$. Each half-integer is thought as a spin, i.e. the label of the irreducible representation of the Lie group $\SU(2)$ of dimension $d_{j}=2j+1$. The dressing of the triangulation with algebraic data from the representation theory of $\SU(2)$ allows to dress each tetrahedron with the 6j recoupling symbol involving the spins attached to its six edges, as illustrated on fig.\ref{fig:tetrahedron}. This leads to the Ponzano-Regge state-sum, or partition function, for a triangulation:
\be
Z^{PR}_{\Delta}
=
\sum_{\{j_e\}}\prod_{e} (-1)^{2j_{e}} d_{je} 
\prod_{t} (-1)^{j^{(t)}_{1}+j^{(t)}_{2}+j^{(t)}_{3}}
 \prod_T \begin{Bmatrix}
j_{1}^{(T)} &j_{2}^{(T)} & j_{3}^{(T)}\\
j_{4}^{(T)} &j_{5}^{(T)} & j_{6}^{(T)}
\end{Bmatrix}
\,,
\ee
involving products of local amplitudes associated to each edge, each triangle and each tetrrahedron.
Due to the Biedenharn-Elliott identity satisfied by the 6j-symbols, this sum does not depend on the details of the triangulation, but only on its overall topology, making it a topological invariant.
The state-sum can be restricted to integer spins, i.e. representations of $\SO(3)\sim\SU(2)/\Z_{2}$, with affecting its topological invariance. This restriction  drops all the odd sign factors from the formula above.
The extension of non-vanishing cosmological constat is achieved by considering the $q$-deformed quantum group $\cU_{q}[{\SU(2)}]$, as realized by the Turaev-Viro model \cite{Turaev:1992hq} (see also the recent works \cite{Dupuis:2013haa,Dupuis:2014fya,Bonzom:2022bpv}). The extension of the model to Lorentzian signature (-++) was later developed in \cite{Barrett_1994,Freidel:2000uq,Davids_2000,davids2001state}, by dressing 3d triangulations with $\SU(1,1)$ representations instead of $\SU(2)$. This leads to a discrete spectrum of time-like intervals and a continuous spectrum of space-like distances \cite{Freidel:2002hx}.

The insertion of matter as topological defects in the Ponzano-Regge model was realized in  \cite{Freidel:2005bb,Freidel:2005me}, by drawing the Feynman diagrams directly on the triangulation.
First, in 3d gravity, mass is a source of curvature, while spin is a source of torsion. Putting spin aside, a point-like particle with mass creates a conical curvature. The mass is given by the angle deficit of the defect, while the particle momentum is measured by the holonomy around the defact.
Second, one must distinguish the bare mass from the physical -or renormalized- mass, which fully takes into account  gravitational effects. The bare mass is directly given by the defect angle $\vphi\in[0,\pi]$ in Planck mass unit, while the physical mass is given by the sine of the angle:
\be
m_{bare}=\ka \vphi\,,\qquad
M_{phys}=\ka\sin\vphi\,,
\ee
which automatically satisfies the Planck mass bound $M\le \ka$. 

For scalar fields, we draw Feynman diagrams on the triangulated space-time by inserting off-shell particles along the triangulation edges.
%a massive particle is thus inserted along an edge as topological defect of angle  $\varphi$.
So, drawing a Feynmann diagram on a subgraph $\Gamma$ in the triangulation, with masse angles $\vphi_{e}$ on the graph edges  gives the following Ponzano-Regge amplitude,
\be
Z^{PR}_{\Delta}[\Gamma]
=
\sum_{\{j_e\}}
\prod_{e \notin \Gamma} d_{j_{e}} \prod_{e \in \Gamma} K_{\varphi_e}(j_e)
\prod_{t} (-1)^{j^{(t)}_{1}+j^{(t)}_{2}+j^{(t)}_{3}}
 \prod_T \begin{Bmatrix}
j_{1}^{(T)} &j_{2}^{(T)} & j_{3}^{(T)}\\
j_{4}^{(T)} &j_{5}^{(T)} & j_{6}^{(T)}
\end{Bmatrix}
\,,
\ee
where $K_{\theta_e}$ corresponds the insertion of Feynman propagator. 
Fields with spins can also be taken into account \cite{Freidel:2004vi,Freidel:2005bb}.
Actually, this formula was previously only written for $\SO(3)$ representations, in \cite{Freidel:2005me,Livine:2008sw}. So we will work out the general case, of $\SU(2)$ representations, in the present work, and notice ambiguities in the definition of the deformed Feynman propagator, which were missed by earlier works.

These quantum gravity amplitudes with (off-shell) matter insertions can be summed exactly and yield  Feynman diagrams of a braided NCQFT, where the 3d Poincar\'e symmetry is upgraded to the Drinfeld double of $\cD[\SU(2)]$.
The effective action of the noncommutative scalar field is defined on the non-commutative space-time $\mathbb{R}_\kappa^{3}$, which is the $\mathbb{R}_3$ space equipped with a non-trivial $\star$-product rendering the coordinates non-commutative with $\su(2)$ commutation relations (or $\su(1,1)$ in the Lorentzian case). The momentum space is then the Lie group $\SU(2)$ (or $\SU(1,1)$ in Lorentzian signature). This makes possible to write the NCQFT action as a group field theory  (see \cite{Oriti:2009wn} for a review of the GFT framework) and actually derive it directly from Boulatov's group field theory for the Ponzano-Regge model \cite{Fairbairn_2007} (see \cite{Boulatov:1992vp} for Boulatov's group field theory paper).

%When the Ponzano Regge model is coupled with a matter field, it has been shown in \cite{Freidel:2005bb} \cite{Freidel:2005me} that gravitational degree of freedom can be integrated out and the result can be described with a noncommutative effective quantum field theory characterized by relations \ref{eq:ncrelations}  between coordinates.
%More precisely, the action $S[\phi]$ of the noncommutative scalar field can be expressed on $\mathbb{R}_3$ 
%equipped with a noncommutative product, called $\star$ product. There are several noncommutative products associated with \ref{eq:ncrelations}, the one study here is of Voros type \cite{Dupuis_2012},  (for discussion about other star product based on the bracket above see \cite{Galluccio_2008}), we denote this space $\mathbb{R}_\kappa^{3}$. The corresponding momentum space became also deformed and can be express with a Lie group based on the Lie algebra $\mathfrak{su}(2)$. The resulting action can be link with Group Field Theory (GFT) as shown in \cite{Fairbairn_2007}. In the Euclidean case, the momentum Lie groups are SO(3) or SU(2), and in the Lorentzian case, SO(2,1) or SU(1,1). The fact that several Lie groups are candidates for encoded momentum space is one ambiguity in the definition of an effective field theory as a group field theory. A study of unitarity would, therefore, give arguments on the group choice.\\

\bigskip

The question of unitarity of this NCQFT was mentioned in \cite{Freidel:2005me} 18 years ago and briefly investigated in \cite{Sasai:2009jm}, and, despite its importance, has been shelved until now.
In fact, it is a fundamental question for noncommutative quantum field theory and quantum gravity.
Indeed, we expect two different results, depending on how we look at the theory. On the one hand, if we consider this model as an effective field theory, derived from a more fundamental theory after integrating out degrees of freedom, unitarity should naturally fail as soon as we reach high enough energy to excite the integrated-out modes. On the other hand, if we view those NCQFTs as legitimate well-defined and mathematically rigorous field theories directly built from scratch on the non-commutative space-time $\mathbb{R}_\kappa^{3}$, is is natural to expect unitarity, the existence of a probability current and conservation of information.

In the present work, we check the unitary at the one-loop level using the Cutkosky rule in both Euclidean and Lorentzian cases. We find that Lorentzian and Euclidean versions of the Cutkosky rule are valid for $\SO(3)$ and $\SO(2,1)$ only for small mass. Furthermore, they are valid for $\SU(2)$ and $\SU(1,1)$ for all masses, if we add an extra term to the Feynman propagator.
This extra term is a 0-mode, actually corresponding to a infinite-dimensional representation of $\SU(2)$ in the Euclidean theory and corresponding to a representation of $\SU(1,1)$ on the light cone, with vanishing Plancherel measure, and usually refered to as  the \textit{limit of discrete series} \cite{knapp2001representation}.
The paper is organized as follows. The first section presents the noncommutative theory for matter coupled to 3d quantum gravity. The second section reviews the Cutkosky rule for standard QFT in commutative space-time. We take special care to clearly derive the  counterpart in Euclidean space-time signature of the optical theorem at one-loop. We then move on to the investigation of the Euclidean noncommutative theory based on $\SO(3)$ and $\SU(2)$. We  describe the different propagators and then compute explicitly the on-shell and off-shell one-loop amplitudes to check the validity of the Cutkosky rule. We underline the key differences between $\SO(3)$ and its double cover.  We then tackle the heart of our work and analyze the Lorentzian theory, for which we also derive propagators and explicitly compute  the one-loop amplitudes for the $\SO(2,1)$ and $\SU(1,1)$ momentum spaces. Finally, we discuss the implication of our results for the unitarity of the NCQFTs and underline interesting lines of future investigation.

%%%%%%%%
\section{Noncommutative scalar field theory on $\R^{3}_{\ka}$}
%\section{Effective noncommutative scalar field theory.}
%%%%%%%%

We are interesting in a class of noncommutative field theory built from curved momentum spaces. Assuming that the momentum space has the structure of a semi-simple Lie group $G$, we consider the following scalar field actions:
\be
S[\phi]
\,=\,
\f12\int_{G} \rd g\,\cK(g)\phi(g)\phi(g^{-1})
+\sum_{n}\f{\lambda_{n}}{n}\int [\rd g]^{n}\,
\delta(g_{1}..g_{n})\,
%\delta\left(\prod_{k=1}^{n}g_{k}\right)\,
\phi(g_{1})..\phi(g_{n})
\,,
\ee
where we integrate using the Haar measure over the Lie group. The kernel $\cK(g)$ defines the kinetic term, while $\delta(g_{1}..g_{n})$ enforces the conservation of momentum in the $n$-valent interaction term. A non-abelian group product translates into a noncommutative addition of momenta. The noncommutativity of spacetime coordinates can then be derived by taking the group Fourier transform of this action (see e.g. \cite{Freidel:2005ec,majid2000foundations,Girelli:2022foc}).

Here, in the context of 3d quantum gravity, we are interested in the case that the Lie group is $G = \SO(3)$ or $\SU(2)$ for an Euclideanspace-time signature and  $G =\SO(2,1)$ or $\SU(1,1)$ for the  Lorentzian theory.
Then the kinetic kernel $\cK$ is taken of the form $p^{2}-m^{2}$.
We will restrict to the cubic self-interaction for the sake of simplicity, in which case the action reads
%
%It has been shown in \cite{Freidel:2005me} that one can derive an effective scalar field theory for matter coupled to 3d quantum gravity in the Ponzano Regge spinfoam model. This effective field theory can also be expressed in the context of group field theory \cite{Fairbairn_2007}, and the resulting action reads: 
%
\begin{equation}
\label{actionG}
S[\phi]
=
\frac{1}{2} \int_G \rd g \; (P^{2}(g) - M^{2}) \phi(g) \phi(g^{-1}) + \frac{\lambda}{3} \int_G [\rd g]^{3} \; \delta(g_1 g_2 g_3) \phi(g_1)\phi(g_2)\phi(g_3) 
\end{equation}
%Where integral are over a general Lie group $G$. In the rest of the paper we consider $G = SO(3)$ or $SU(2)$ for Euclidean theory of 3d quantum gravity and  $G = SO(2,1)$ or $SU(1,1)$ for Lorentzian theory of 3d quantum gravity.
%
The momentum $P(g)$ is defined as the projection on the group element $g$ on the Euclidean or Lorentzian Pauli matrices depending on the chosen space-time signature. More precisely,
\begin{itemize}
\item In the Euclidean theory:
\\
We project on the $\su(2)$ Pauli matrices $(\sigma_0,\sigma_1, \sigma_2)$, satisfying  $(\sigma_{a})^{2}=\id$ for all $a=0,1,2$.
In the fundamental two-dimensional representation, a $\SU(2)$ group element decomposes onto the Pauli matrices and the identity as $g = \cos\theta\, \id + i \sin\theta\, \hu \cdot \vec{\sigma}$, with $\theta \in [0,2\pi]$ the half-angle of rotation and $\hu \in \cS^{2}$ the rotation axis.
Obviously this is a redundant parametrization, with the identification of the group elements $(\theta,\hu)$ and ($2\pi-\theta$,$-\hu$), so we can safely restrict to angles $\theta \in [0,\pi]$.
The projection is simply defined as:
\be
\vec{P}(g)
=
\f\kappa{2i}  \tr(g \vec{\sigma})
=
\kappa \sin\theta \,\hu\,,
\qquad
P^{2}(g) = \kappa^{2} \sin^{2}\theta
\leq \kappa^{2}
\,.
\ee
This is a well-defined momentum map, since it is invariant under the exchange $(\theta,\hu)\leftrightarrow(-\theta,-\hu)$.
%
%The momentum norm is bounded $0 \leq P^{2} $, due to the compactness of the group.
%
It is crucial to notice that the map $g\in\SU(2)\mapsto \vP(g)\in\R^{3}$ is onto but not bijective. Indeed changing the sign of the cosine, i.e. mapping the angle $\theta\rightarrow \pi-\theta$, does not change the momentum:
\be
\vP(\theta,\hu)=\vP(\pi-\theta,\hu)
\,.
\ee
This two-fold degeneracy can be lifted by using the group $\SO(3)$ instead of $\SU(2)$.
Indeed, a $\SO(3)=\SU(2)/\Z_{2}$ group element is defined by the equivalence relation $g\sim -g$, which corresponds to $(\theta,\hu)\sim (\pi-\theta,-\hu)\sim(\theta+\pi,\hu)$ in the angle-axis parametrization. We can therefore choose $\SO(3)$ group elements as corresponding to angles $\theta\in[0,\pi/2]$, thus with positive sine and cosine.
Then a 3-vector $\vP\in\R^{3}$, norm-bounded by $|\vP|\le \ka$, corresponds to a unique $\SO(3)$ group element with $\hu=\vP/|\vP|$ (thus pointing in the same direction) and $\sin\theta=|\vP|/ \ka\ge 0$. 

\item In the Lorentzian theory:
\\
We work with space-time signature $\eta := (+, -,-)$ and we use the $\su(1,1)$ Pauli matrices $\tau_0 = \sigma_0$, $\tau_{1,2} = i \sigma_{1,2}$.
Group elements of $\SU(1,1)$ are of three types.
Elliptic or time-like group elements read $g= \cos\theta \id+ i \sin\theta \, \hu \cdot \vec{\tau}$, with $\hu^{2} = + 1$ and the class angle $\theta \in [0,2\pi]$.
Similarly to the Euclidean case, this parametrization is redundant and $g(\theta,\hu)=g(2\pi-\theta,-\hu)$. If we want to keep a definite direction for the rotation axis $\hu$, we should lift this degeneracy and restrict to angles $\theta\in[0,\pi]$. This allows to identify elements with $u_0 > 0$ as positive time-like, and those with $u_0 < 0$ as negative time-like.
%
%Unlike the Euclidean situation, the SO(2,1) element's time orientation makes only sense if we restrict the rotation angle $\theta \in [0,\pi]$.
%
Hyperbolic or space-like group elements are $g = \pm (\cosh t + i \sinh t \, \hat{a} \cdot \vec{\tau})$, with $\hat{a}^{2} = -1$. Note that the sign $\pm$ can not be absorbed by a modification of $t$ or $\ha$. We have the same degeneracy as for elliptic group elements with $g(t,\ha)=g(-t,-\ha)$, which is simply lifted by restricting to positive boost parameters $t\ge 0$.
Finally, null group elements are $g = \pm Id + i \vec{l}\cdot\vec{\tau}$, with $\vec{l}^{2} = 0$.

Momentum is defined, as in the Euclidean case, by projecting group elements on the Pauli matrices,
\be
\vec{P}(g)
=
\f\kappa{2i}  \tr(g \vec{\tau})
=
\left\{
\begin{array}{l}
\kappa \sin\theta \,\hu
\,,\vspace{1mm}\\
\kappa \sinh t \,\hat{a}
\,,\vspace{1mm}\\
\kappa \vec{l}
\,,
\end{array}
\right.
\ee
where only elliptic group elements have bounded momenta.
%
%Momentum for elliptic elements is similar to the Euclidean one, $\vec{P}(g) = \kappa \sin\theta \hu$, those elements describe massive particles and have their momentum bounded $P^{2} \leq \kappa^{2}$. For hyperbolic elements, we have $\vec{P}(g) = \kappa \sinh t \hat{a}$. This time, the gauge group SO(2,1) is no more compact, and the momentum is unbounded on the hyperbolic sector.

The map $g\in\SU(1,1)\mapsto\vP(g)\in\R^{1,2}$ is once again onto but not bijective. This can be lifted by using the subgroup $\SO(2,1)=\SU(1,1)/\Z_{2}$, defined by the equivalence relation $g\sim-g$. This amounts to restricting the class angle of elliptic elements $\theta\in[0,\pi/2]$, dropping the sign $\pm$ in front of hyperbolic and null elements. This makes the momentum map bijective.

\end{itemize}

At this stage, in both Euclidean and Lorenztian signatures, we clearly see the difference between the bare mass $m=\ka\theta$ and the renormalized mass $M=\ka\sin\theta$. The bare mass $m$ keeps increasing as the defect angle $\theta$ grows from 0 to $\pi$, while the renormalized mass is not in one-to-one correspondance with the angle: it increases in the range $\theta\in[0,\pi/2]$, reaches its maximum at $\theta=\pi/2$ then decreases. We therefore expect the large mass sector $\theta\in[\pi/2,\pi]$ to exhibit large (quantum) gravitational effects.
We will indeed see that the field theories based on $\SO(3)$ or $\SO(2,1)$ momentum space will fail to be unitary due to this large mass sector. In order to correct this bad feature, one has to move to the double-covers $\SU(2)$ or $\SU(1,1)$, and introduce a Feynman propagator that distinguishes the angles $\theta$ and $(\pi-\theta)$. So instead of merely using $\sin\theta$, we will also simply use $\cos\theta$ to define the kinematics of the NCQFT. At the end of the day, this will fix the unitarity (at least, at the one-loop level).

\medskip

Using this momentum $\vP(g)$, one can perform a change of variables, from group elements $g$ to usual vectors $\vP$, and write our field theory \eqref{actionG} in term of standard momenta variables \cite{Freidel:2005me}:
\beq
\label{actionG1}
S[\phi]
&=&
\frac{1}{2} \int_{\R^{3}} \mu(\vP)\rd^{3}\vP \, \,\Big{(}\vP^{2}(g) - M^{2}\Big{)}\, \phi(+\vP) \phi(-\vP)
\\
&+& 
\frac{\lambda}{3} \int \prod_{k=1}^{3}\mu(\vP_{k})\,\rd^{3}\vP_{k} \,\,
\delta\Big{(}\vP_1\oplus \vP_2 \oplus \vP_3\Big{)}\,
\phi(\vP_1)\phi(\vP_2)\phi(\vP_3)
\,
\nn
\eeq
where the non-commutative addition of momenta  is inherited from the non-abelian group multiplication,
\be
\vP(g_{1})\oplus \vP(g_{2})=\vP(g_{1}g_{2})\,.
\ee
The measure factor $\mu(\vP)$ is inherited from the Haar measure. It reads explicitly for $G=\SO(3)$ or $\SO(2,1)$:
\begin{equation}
    \int_G \rd g =  \frac{1}{\pi^{2} \kappa^{3}} \int_{\R^{3}} \frac{\rd^{3} \vec{P}}{\sqrt{1- \frac{P^{2}}{\kappa^{2}}}}\,.
\end{equation}
Remember that momenta are always bounded, $|\vec{P}| \leq \kappa $, for $G=\SO(3)$, while this bound also holds for time-like vectors in $\R^{2,1}$ (i.e. elliptic group elements) for $G=\SO(2,1)$.
For the $\SU(2)$ or $\SU(1,1)$ groups, the change of variables $g\mapsto \vP(g)$ is not bijective, so we need to sum over an extra sign when writing the theory in terms of vector momentum. Details on the resulting measures can be found in the appendix \ref{appendixA}.

In the limit of an infinite Planck mass, when the deformation parameter is sent to infinity, $\kappa \to \infty$, we recover the standard commutative  momentum space $\mathbb{R}^{3}$ equipped with the  trivial addition of momenta,
\begin{equation}
S_{\text{undeformed}}[\phi]
=
\frac{1}{2} \int \rd \vec{p}\,{}^{3} (p^{2} - m^{2}) \phi(+\vec{p}) \phi(-\vec{p})
+
\frac{\lambda}{3} \int [d^{3} \vec{p}]^{3} \delta(\vec{p}_1 + \vec{p}_2 + \vec{p}_3) \phi(\vec{p}_1)\phi(\vec{p}_2)\phi(\vec{p}_3)
\,.
\end{equation}
One can then compute the quantum gravity corrections to standard quantum field theory order by order in $\ka$, as shown in \cite{Freidel:2005bb}. This formalism clarifies how the Planck mass $\ka$ encodes the quantum gravity fluctuations and contributions to the effective field theory.

%Let us underline that this Euclidean theory is the $G_N \to 0 $ limit of a scalar field coupled to 3d Euclidean quantum gravity with a metric signature $(+,+,+)$. It is unrelated prior to the Lorentzian theory through a Wick rotation. \footnote{In fact we have the following transformation for coordinates $x_0 \to x_0$ and $x_j \to i x_j$} In particular, the kinetic term here is $(p^{2} - m^{2})$ instead of usual Euclidean $(p^{2} + m^{2}) $. We see this Euclidean model as a toy model, mathematically simpler than its Lorentzian counterpart, based on the non-compact group $SO(2,1)$ instead of $SO(3)$.

%There exists between this noncommutative space and SO(3) a Fourier transform \footnote{Explicit expression of the Fourier transform and its inverse reads:  \begin{equation*} \varphi(x) =  \int_{\SO(3)} dg \; \Tilde{\varphi}(g) \exp^{\frac{1}{2\kappa} \tr(xg)} \end{equation*} \begin{equation*}\Tilde{\varphi}(g) = \int_{\R^{3}}  \frac{d^{3} x}{8 \pi \kappa^{3}} \varphi(x) \star \exp^{\frac{1}{2 \kappa}\tr(xg^{-1})}\end{equation*}}.
%
%In coordinate space the action \ref{actionG1} gives rise to a noncommutative field theory with deformation parameter $\kappa$ which encode the quantum gravity fluctuation of the effective field theory.

It is possible to define a group Fourier transform to go from the action \eqref{actionG} in terms of group variables, or the action \eqref{actionG1} in terms of noncommutative momenta, to an action written in coordinate space \cite{Freidel:2005me,Freidel:2005ec}. The key is that the group product, or equivalently the noncommutative addition of momenta, equips functions over the 3d coordinates a non-trivial $\star$-product, which is simply defined in terms of plane-wave multiplication on $\R^{3}$:
\begin{equation}
\forall X\in\su(2)\,,\,\,
e^{\frac{1}{2 \kappa} \tr(X g_1) } \star e^{\frac{1}{2 \kappa} \tr(X g_2) } = e^{\frac{1}{2 \kappa} \tr(X g_1 g_2) }
\,,
\quad\textrm{and}\qquad
\forall \vx\in\R^{3}\,,\,\,    
e^{i\vx\cdot\vP_{1} } \star e^{i\vx\cdot\vP_{2}} = e^{i\vx\cdot(\vP_{1}\oplus\vP_{2})}
\,.
\end{equation}
This $\star$-product leads to a non-commutativity of the space-time coordinates,
\begin{equation}
[x_i, x_j] = i \kappa^{-1} \epsilon_{ij}{}^{k} x_k 
\,,\quad
[x_i,P_j] = i \sqrt{1 - \kappa^{-2} P^{2}} \delta_{ij} - i \kappa^{-1} \epsilon_{ij}{}^{k} P_k
\,.
\end{equation}
This allows to write the field action in coordinate space \cite{Freidel:2005bb}:
\begin{equation}
S[\phi]
=
\int \frac{\rd^{3} x}{8 \pi \kappa^{3}}  \, 
\left[
-\frac{1}{2} \partial_\mu \phi \star \partial^{\mu} \phi(x) - \frac{1}{2}M^{2} \phi \star \phi (x) + \frac{\lambda}{3} (\phi \star \phi \star \phi(x)
\right]
\,.
\end{equation}
For more details on this $\star$-product, the interested reader can check \cite{Freidel:2005ec,Livine:2008sw,Livine:2008hz,Dupuis_2012}.
The aim of the present work is to study the unitarity of these noncommutative field theories in both Euclidean and Lorentzian signatures.

%%%%%%%%
\section{One-loop integrals and cut rule}
%\section{Cutkoksy rule for one-loop in 3d scalar commutative quantum field theory.}
%%%%%%%%

\begin{figure}[h!]
\includegraphics[width=100mm]{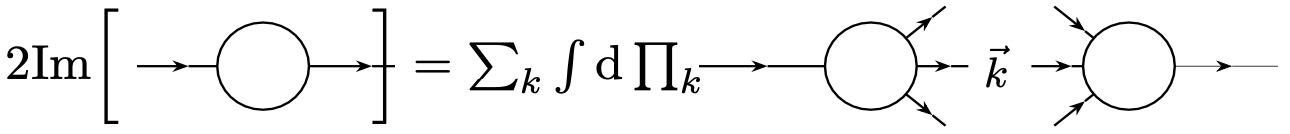}
 \caption{{On the left side the imaginary part of the amplitude (off-shell) and on the right the amplitude decomposed over all possible intermediate states (on-shell). }\label{bellefigure}}
\end{figure}

In this section, we compute the leading one-loop correction to the field propagator for the standard undeformed Euclidean and Lorentzian 3d theories. Unlike the 4d theory, where one-loop amplitudes have logarithmic divergences, we have finite expressions in 3d.
We can thus check unitarity with the Cutkosky rule at one loop without any regularization.  In a Lorentzian space-time, for time-like external momentum, this rule states that the imaginary part of the off-shell one-loop amplitude is equal to its on-shell counterpart, as illustrated on figure \ref{bellefigure}.
If we denote $L_m^{L}(\vec{p})$ the amplitude for Lorentzian off-shell one loop, with external momentum $\vec{p}$, and $L_m^{L,0}(\vec{p})$ the Lorentzian on-shell one loop, they satisfy the following relation:
\begin{equation}
    i\; \text{Im}(L_m^{L}(\vec{p})) = L_m^{L,0}(\vec{p})
    \,.
\end{equation}
We review the computation these amplitudes below and check this relation.

We also derive a Euclidean version of this cut rule, relating the Euclidean off-shell and on-shell one-loop amplitudes. We will then be ready to check if the same Euclidean and Lorentzian rules are also satisfied by the non-commutative quantum field theory at one-loop.

%%%
\subsection{Field theory in Euclidean signature}
%%%

The Feynman propagator for mass $m$ in  Euclidean signature is the solution of the Klein-Gordon equation  with Dirac-$\delta$ source $(\Delta + m^{2})G_m^{\rE}(\vec{x}) = - \delta^{3}(\vec{x})$. It is given by the Fourier transform of $K_m^{\rE}(\vec{p}) = (p^{2} - m^{2} + i \epsilon)^{-1}$, where $\epsilon > 0$ a regulator,
\begin{equation}
    G_m^{\rE}(\vec{x}) = \int \frac{d^{3} \vec{p}}{(2 \pi)^{3}} \frac{e^{i \vec{x}\vec{p}}}{p^{2} - m^{2} + i \epsilon}
    \,,
\end{equation}
%
%\begin{equation*}
%    \frac{1}{p^{2} - m^{2} + i \epsilon} = \frac{1}{4 \pi} \int d^{3} \vec{x} \frac{e^{-i(m-i \epsilon)|\vec{x}|}}{|\vec{x}|}e^{-i \vec{x} \vec{p}}
%\end{equation*}
%Explicitly computing the Fourier transform we can prove that sing the following spherical integration formula:$\int_{\cS} d^{2} \hat{x} e^{i \hat{x} \hat{u}} = 4 \pi \frac{\sin(|\hat{u}|)}{|\hat{u}|}$ it is straightforward to show that:
%
%\begin{equation*}
%    \frac{1}{p^{2} - m^{2} + i \epsilon} = \frac{1}{4 \pi} \int d^{3} \vec{x} \frac{e^{-i(m-i \epsilon)|\vec{x}|}}{|\vec{x}|}e^{-i \vec{x} \vec{p}}
%\end{equation*}:
which is straightforward to integrate,
\begin{equation}
    G_m^{\rE}(\vec{x}) = \frac{e^{-i (m-i \epsilon)|\vec{x}|}}{4 \pi |\vec{x}|}
    \,.
\end{equation}
The one-loop diagram with two momentum insertions gives the  leading order correction to the field propagator. Its amplitude is an integral over the momentum running around the loop:
\begin{equation}
    L_m^{\rE}(\vec{p}) = i \int \frac{d^3 \vec{q} }{(2 \pi)^3} K_m^{\rE}(\vec{q}) K_m^{\rE}(\vec{ p} - \vec{q}) = \frac{i}{4 \pi |\vec{p}|} \int_0^{\infty} dr \frac{e^{-2i mr}}{r} \sin{|\vec{p}|r}\,,
\end{equation}
which can be computed exactly, yielding:
\begin{equation}
        L_m^{\rE}(\vec{p}) = \left\{
        \begin{array}{ll}
             \displaystyle\frac{-1}{8 \pi |\vec{p}|}  \log\left(\frac{2m - |\vec{p}|}{2m + |\vec{p}|}\right)\,, & \quad \text{if } |\vec{p}| < 2m\,, \\
            \displaystyle \frac{- 1}{8 \pi |\vec{p}|} \log\left(\frac{2m - |\vec{p}|}{2m + |\vec{p}|}\right)  +  \frac{i}{8 |\vec{p}|}\,, & \quad \text{if }  |\vec{p}| > 2m\,.
        \end{array}
    \right.
\end{equation}
One notices the apparition of a non-vanishing imaginary part above the two-particle threshold $|\vec{p}| > 2m$.
Let us compare this exact one-loop diagram to the corresponding on-shell integral:
\begin{equation}
    L_m^{0,\rE}(\vec{p})  = \frac{i}{2} \int \frac{d^{3} \vec{q}}{(2 \pi)^{3} } (2 \pi) \delta(|\vec{q}|^{2} - m^{2})( 2 \pi ) \delta(|\vec{q} + \vec{p}|^{2} - m^{2}) = \left\{
    \begin{array}{ll}
        \displaystyle\frac{i}{8 |\vec{p}|} & \text{if } |\vec{p}| < 2m \\
        0 & \text{if } |\vec{p}| > 2m 
    \end{array}
\right.
\end{equation}
\begin{figure}[htb!]
\includegraphics[width=50mm]{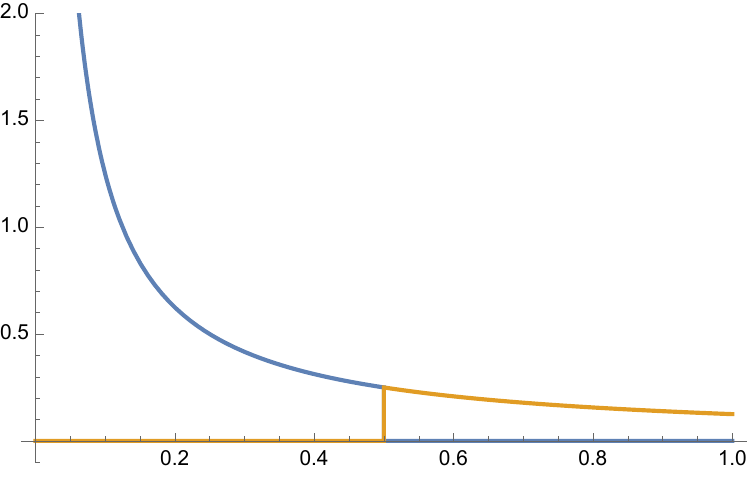}
\caption{\label{plot:unitaryE}
Plots for  $\text{Im}\,L^{\rE}_{m}(\vec{p})$, in \textcolor{red}{red}, and $-iL^{0, \rE}_{m}(\vec{p})$, in \textcolor{blue}{blue}, in terms of $|p|/m$: they have disjoint domains and always sum up to $1/8|p|$.
}
\end{figure}
We find that the imaginary part of the one-loop diagram is  equal to its on-shell counterpart up to an awkward inversion of the mass-momentum threshold condition $|p|>< 2m$. This means that, eventhough this Euclidean theory is clearly not unitary,  we nevertheless have the following relation between the imaginary part of the off-shell  and on-shell one-loop integrals, as plotted on fig.\ref{plot:unitaryE}:
\be
    i \,\text{Im}(L^{\rE}_{m}(\vec{p})) +  L^{0, \rE}_{m}(\vec{p})
    = \frac{i}{8 |\vec{p}|}
    \,,\quad\forall \vec{p}
    \,,
\ee

with either functions vanishing when the other does not:  $i \,\text{Im}(L^{\rE}_{m}(\vec{p})$ vanishes for $|p|<2m$, while $L^{0, \rE}_{m}(\vec{p})$ vanishes for $|p|>2m$. In some sense, one could say that these two functions complete each other.
This is our Euclidean counterpart of the Lorentzian optical theorem. We will seek to check whether a similar relation holds for the  Euclidean noncommutative field theory.

%%%
\subsection{Field theory in Lorentzian signature}
%%%

\begin{figure}[h!]
\begin{subfigure}[t]{0.28\linewidth}
%\begin{tikzpicture}[scale=.3]
%    \draw (0,7.6) ellipse (2.08cm and .3cm);
%    \draw (0,-7.6) ellipse (2.08cm and .3cm);
%
%
%\coordinate(A1) at (-3.8,7.8);
%\coordinate(A2) at (3.8,7.8);
%\coordinate(B1) at (-3.8,-7.8);
%\coordinate(B2) at (3.8,-7.8);
%\coordinate(C1) at (0,1.6);
%\coordinate(C2) at (1.8,8);
%\coordinate(D1) at (0,-1.6);
%\coordinate(D2) at (-1.8,-8);
%    
%    \draw [->, blue, thick] (C1) -- (C2);
%    \draw [->, blue, thick] (D1) -- (D2);
%    \draw (A1) parabola bend (0,2) (A2);
%    \draw (B1) parabola bend (0,-2) (B2);%\draw (A1) arc (A2);   
%    %\addplot [domain=-\dom:\dom] ({\a*cosh(\x)},{\b*sinh(\x)});
%    %\addplot [domain=-\dom:\dom] ({-\a*cosh(\x)},{\b*sinh(\x)});
%
%\end{tikzpicture}
%
\includegraphics[width=30mm]{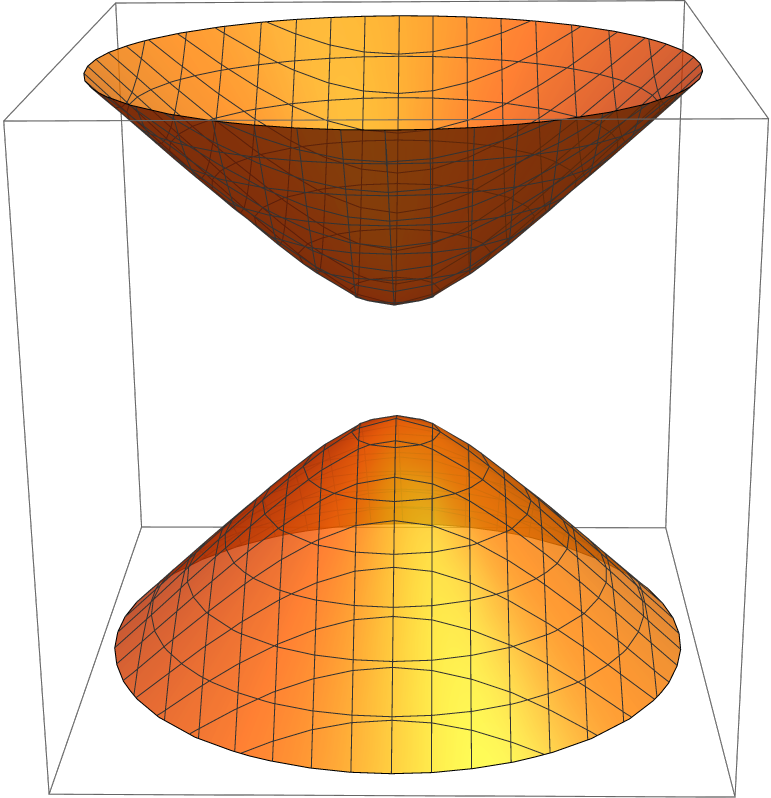}
\subcaption{
Two-sheet hyperboloid of normed time-like momenta, $\cH^{+}$ and $\cH^{-}$.}
\end{subfigure}
\hspace{2mm}
\begin{subfigure}[t]{0.28\linewidth}
%\begin{tikzpicture}[scale=.3]
%    %\draw (0,0) ellipse (.55cm and .2cm);
%    \draw (0,7.6) ellipse (2.08cm and .3cm);
%    \draw (0,-7.6) ellipse (2.08cm and .3cm);
%
%\coordinate(A1) at (-4,8);
%\coordinate(A2) at (4,8);
%\coordinate(B1) at (-4,-8);
%\coordinate(B2) at (4,-8);
%\coordinate(C1) at (0,0);
%\coordinate(C2) at (3.5,8);
%
%
%    \draw [->, blue, thick] (C1) -- (C2);
%    
%    \draw (A1)--(B2);
%    \draw (A2)--(B1);
%
%\end{tikzpicture}
%
\includegraphics[width=30mm]{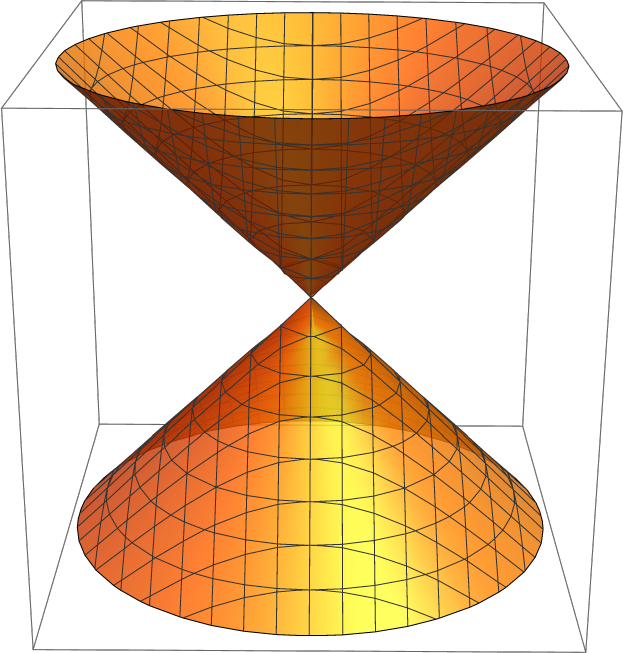}
\subcaption{
Light cone of null 3d momenta.
}

\end{subfigure}
\hspace{2mm}
\begin{subfigure}[t]{0.28\linewidth}
%\begin{tikzpicture}[scale=.7]
%\def\xm{5}
%\def\ym{10}
%\def\df{3}
%\def\dom{2}
%
%%\def\ecc{1.44022}
%\def\ecc{2.3}
%\def\a{1}
%\def\b{(\a*sqrt((\ecc)^2-1)} 
%
%\begin{axis}[scale=.8,
%    hide axis,
%    xmin=-\xm,xmax=\xm,
%    ymin=-\ym,ymax=\ym]
%
%\coordinate(D1) at (0.8,0);
%\coordinate(D2) at (4,4);
%
%    
%    \draw [->, blue, thick] (D1) -- (D2);
%    \addplot [domain=-\dom:\dom] ({\a*cosh(\x)},{\b*sinh(\x)});
%    \addplot [domain=-\dom:\dom] ({-\a*cosh(\x)},{\b*sinh(\x)});
%    
%\end{axis}
%\end{tikzpicture}
%
\includegraphics[width=30mm]{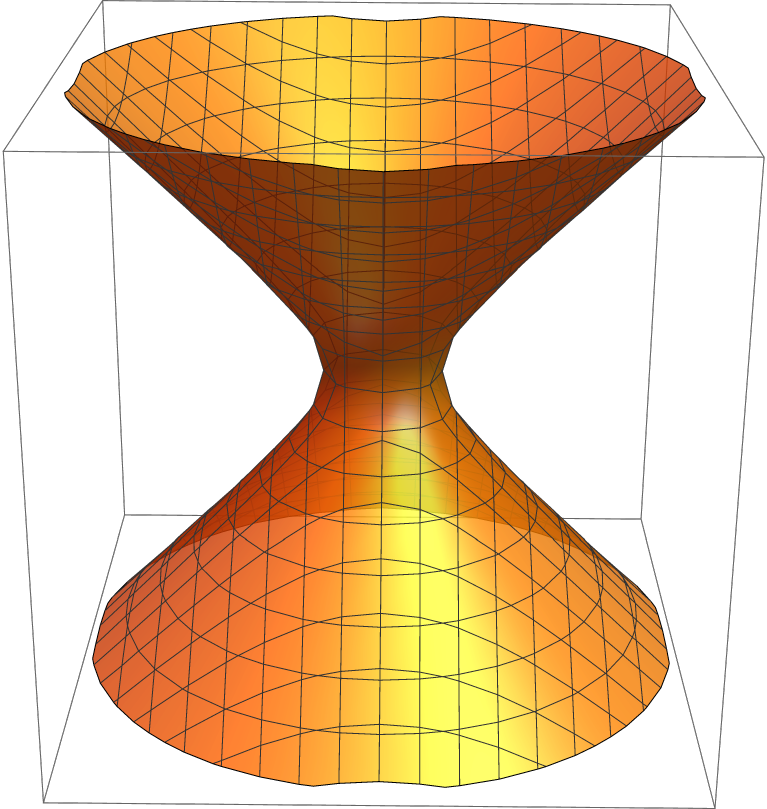}
\subcaption{
One-sheet hyperboloid of norm space-like 3d momenta.
}
\end{subfigure}

\caption{\label{fig:hyperbola}}
\end{figure}

Let us now check how the one-loop calculation works in the more standard Lorentzian field theory, with metric signature $(+,-,-)$. There are three types of momenta, as illustrated on fig.\ref{fig:hyperbola}. There are time-like vectors characterized by a positive norm, $p^{2} > 0$. They live in one of the two sheet hyperboloids of the Lorentzian space. We distinguished positive time-like vectors, with positive time component $p_0 >0$, living in the upper hyperboloid $\cH^{+}$ from negative time-like vectors, with $p_0 < 0$, living in the lower hyperboloid $\cH^{-}$. The second one is space-like vectors, which have a negative norm, i.e., $p^{2} < 0$, and they live in the one-sheet hyperboloid.
%
%Those elements are represented on the three figures below \ref{hyperbo}. 

The Feynman propagator for mass $m$  is the solution of the Klein-Gordon equation  with Dirac-$\delta$ source. It is given in momentum space by $K_m^{\rL}(p) = (\vec{p}^{2}-m^{2} + i \epsilon)^{-1}$, where the momentum norm is $\vec{p}^{2} = p_0^{2} - p_1^{2} - p_2^{2}$.
%For time-like momentum $\vec{p}^{2} = |\vec{p}|^{2}$, where $|.|$ is the absolute value , and for space-like momentum $\vec{p}^{2} = -|\vec{p}|^{2}$. 
We write the off-shell one-loop using Feynman's parametrization:
%\footnote{For two complex numbers such that $0$ is not in the segment connecting $A$ to $B$ $\frac{1}{AB}= \int_0^{1} \frac{dx}{(Ax + B(1-x))^{2}}$}
\begin{equation}
L_m^{\rL}(p)
=
i \int \frac{d^{3} q}{(2 \pi)^{3}}K_m^{\rL}(q) K_m^{\rL}(q -p)
=
{i \int \frac{d^{3} q}{(2 \pi)^{3}} \int_0^{1} dx \frac{1}{((q-px)^{2} - m^{2} -p^{2}x(1-x) + i \epsilon)^{2}}}
\,.
\end{equation}
It is possible to compute exactly those integrals, taking special care of properly distinguishing time-like and space-like momenta.
The integral in $q$ can be computed by a Wick rotation and gives:
\begin{equation}
L_m^{\rL}(p)
%=
%i \int \frac{\rd^{3} q}{(2 \pi)^{3}} \int_0^{1} \frac{\rd x}{(q^{2} + p^{2} x(1-x))^{2} - m^{2}+ i \epsilon)^{2}}
=
\int_0^{1} \frac{\rd x}{-8 \pi |p|\sqrt{-p^{2}x(1-x) + m^{2} -i\epsilon}}
%=
%\int_0^{1} \frac{\rd x}{-8 \pi |p|\sqrt{-|p|^{2}x(1-x) + m^{2} -i\epsilon}}
\end{equation}
%After a Wick rotation, we computed the first integral in $q$ and got: $(-8 \pi |p|\sqrt{-|p|^{2}x(1-x) + m^{2} })^{-1}$.
%

Let us first consider an external time-like momentum, with $p^{2}>0$.
% i.e., $p^{2} = |p|^{2}$.
For $|p| <2m$, the interior of the square root is always positive, and we recognize the primitive of $\arcsinh$.
For $|p| > 2m$, we have a branch cut, producing an extra imaginary contribution. Overall, one gets:
\begin{equation}
    L_m^{\rL}(p) = \left\{
        \begin{array}{ll}
             \displaystyle\frac{-1}{8 \pi |p|}  \log\left(\frac{2m - |p|}{2m + |p|}\right)\,, & \quad \text{if }   |p| < 2m\,, \\
             \displaystyle\frac{- 1}{8 \pi |p|} \log\left(\frac{2m - |p|}{2m + |p|}\right)  +  \frac{i}{8 |\vec{p}|}\,, &\quad \text{if }   |p| > 2m \,.
        \end{array}
    \right.
\end{equation}
On the other hand, the computation of the on-shell one-loop amplitude amounts to integrating $\delta$-functions:
\begin{equation}
    L_m^{0, \rL}(p) = \frac{i}{4 \pi} \int \rd^{3} q \delta(q^{2} - m^{2}) \delta((q -p)^{2}- m^{2}) = \left\{
        \begin{array}{ll}
             0 & \text{if }  |p| < 2m \\
               \displaystyle\frac{i}{8 |\vec{p}|} & \text{if }   |p| > 2m 
        \end{array}
    \right.
\end{equation}
As expected, we recover the optical theorem at the one-loop level for time-like elements,
\begin{equation}
    i \; \text{Im}(L^{\rL}_{m}(p)) =  L^{0, \rL}_m(p)
    \,.
\end{equation}
Interestingly, the off-shell one-loop integrals  are the same in Euclidean and Lorentzian theories, whereas the on-shell one-loop integrals have inverse conditions on $|p| <> 2m$. We expect to find a similar behavior for the  noncommutative theories.

For space-like external momentum, $p^{2}<0$,  the off-shell integrals do not have any imaginary part,
%$p^{2} = - |p|^{2}$ and therefore the interior of the square root $\sqrt{|p|^{2}x(1-x) + m^{2}}$ is always positive for $x \in [0,1]$. We did not get any imaginary part for the off-shell one-loops of space-like elements:
\begin{equation}
    L_m^{\rL}(\vec{p}) =  \frac{-1}{4 \pi |p|}  \text{arcsin} \left(\frac{|p|}{\sqrt{4m^{2} + |p|^{2}}} \right)
    \,,
\end{equation}
while the on-shell integral gives:
\begin{equation}
    L_m^{0, \rL}(p) = \frac{i}{4 \pi} \int \rd^{3} q \,\delta(q^{2} - m^{2}) \delta((q -p)^{2}- m^{2}) =  \frac{i}{2}\frac{1}{\sqrt{4 m^{2}+ |p|^{2}}}
    \,.
\end{equation}
The Cutkosky rule unsurprisingly fails in this case. This is to be expected, since space-like momentum us not a physical state, so that one can not expect the optical theorem to hold. Moreover, there is no obvious counterpart of the cut rule for this configuration.

%%This result is not surprising; in fact, the imaginary part of the off-shell one-loop for time-like momentum elements comes from the multi-particle creation threshold $p^{2}> (2m)^{2} $. Space-like momentum did not describe any physical particles; the corresponding associated quantity would be the hypothetical tachyon, which does not have such a threshold because of its negative momentum norm. \\
%%On-shell one-loop for time-like elements reads : \footnote{ We use Lorentzian parametrization of vector, $\vec{q} = q (\sinh(s) \cos(\gamma), \sinh(s) \sin(\gamma), \cosh(s)) $ to compute : \begin{equation}
%%        L_m^{0}(\vec{p})  = \frac{i m}{2} \int_0^{\infty} ds  \sinh{(s)}   \delta( - p^{2} + 2 p m \sinh{s} ) = \frac{im}{2} \int_{0}^{\infty} \frac{Yd Y}{\sqrt{1+Y^{2}}}  \delta(- p^{2} + 2 p m Y) 
%%\end{equation}}
%%\begin{equation}
%%    L_m^{0, \rL}(p) = \frac{i}{4 \pi} \int d^{3} q \delta(q^{2} - m^{2}) \delta((q -p)^{2}- m^{2}) =  \frac{i}{2}\frac{1}{\sqrt{4 m^{2}+ |p|^{2}}}
%%\end{equation}
%%We find that the on-shell one-loop is non-zero and that the Cutkosky rule fails for space-like momentum. The on-shell one is always zero, whereas the off-shell one has no imaginary part. Physically, external particles with negative mass correspond to tachyons; we can conclude that the Cutkosky rule is not verified for such hypothetical particles. Furthermore contrary to the Euclidean version of Cutkosky rule here we do not find any relation between off-shell and on-shell one loops.  

%%%%%%%%
\section{NCQFT Unitarity and Feynman Propagators: Euclidean case}
%\section{Unitarity for noncommutative theories}
%%%%%%%%

In this section, we detail the structure of the one-loop integral for the noncommutative quantum field theory on the Euclidean $\R^3_{\ka}$. In particular, we compute both Hadamard and Feynman propagators. We first tackle the $\SO(3)$ momentum space, then derive work out the case of the $\SU(2)$ momentum space.
As pointed out earlier, the $\SO(3)$ field theory allows for a one-to-one map between $\SO(3)$ group elements and $\R^{3}$ momenta. However, since the Lie group $\SO(3)=\SU(2)/\Z_{2}$ is obtained by a $\Z_{2}$ identification, we do expect an impact of quotienting by this discrete $\Z_{2}$ group. We will explain how it affects unitarity and actually ruins it at high momenta. On the other hand, the $\SU(2)$ field theory has a degenerate two-to-one map between $\SU(2)$ group elements and $\R^{3}$ momenta, so one needs to investigate and understand the physical impact of this degeneracy.

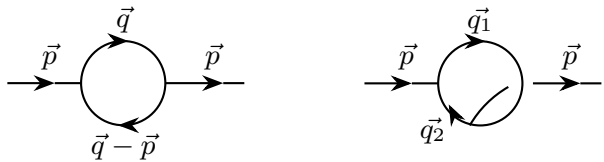
\begin{figure}[htb!]
\centering
\begin{subfigure}[h]{0.2\textwidth}
\begin{tikzpicture}[scale=0.7]
    \coordinate(O) at (-5.3,0);
    \coordinate(5) at (-7.5,0);
    \coordinate(6) at (-6.1,0);
    \coordinate(7) at (-6.6,0);
    \coordinate(52) at (-4.5,0); %11.3
    \coordinate(62) at (-3.4,0);
    \node[scale=1.2] at (-6.7,0.5) { $\vec{p}$};
    \node[scale=1.2] at (-3.6,0.5) { $\vec{p}$};
    
    \draw[ thick,arrows = {-Stealth[fill=black,length=8pt]}](5)--(7);
    \draw[ thick](7)--(6);
    
    \draw[ thick,arrows = {-Stealth[fill=black,length=8pt]}](52)--(62);
    \coordinate(72) at (-3,0);
    \draw[ thick](62)--(72);
    
	\draw[ thick] (O) circle (0.8) ;

    \node[scale=1.2] at (-5.3,1.2) { $\vec{q}$};
    \node[scale=1.2] at (-5.3,-1.2) { $\vec{q} - \vec{p}$};

    \draw[ thick,arrows = {-Stealth[fill=black,length=8pt]}](-5.3,0.8)--(-5.2,0.8);
    \draw[ thick,arrows = {-Stealth[fill=black,length=8pt]}](-5.3,-0.8)--(-5.4,-0.8);
\end{tikzpicture}
\end{subfigure}
\hspace*{10mm}
\begin{subfigure}[h]{0.2\textwidth}
\begin{tikzpicture}[scale=0.7]
    \coordinate(O) at (-5.3,0);
    \coordinate(5) at (-7.5,0);
    \coordinate(6) at (-6.1,0);
    \coordinate(7) at (-6.6,0);
    \coordinate(52) at (-4.3,0); %11.3
    \coordinate(62) at (-3.4,0);
    \node[scale=1.2] at (-6.7,0.5) { $\vec{p}$};
    \node[scale=1.2] at (-3.6,0.5) { $\vec{p}$};
    
    \draw[ thick,arrows = {-Stealth[fill=black,length=8pt]}](5)--(7);
    \draw[ thick](7)--(6);
    
    \draw[ thick,arrows = {-Stealth[fill=black,length=8pt]}](52)--(62);
    \coordinate(72) at (-3,0);
    \draw[ thick](62)--(72);
    
	\draw[ thick] (O) circle (0.8) ;

    \node[scale=1.2] at (-5.3,1.2) { $\vec{q_1}$};
    \node[scale=1.2] at (-6.2,-0.9) { $\vec{q_2}$};

    \draw[ thick,arrows = {-Stealth[fill=black,length=8pt]}](-5.3,0.8)--(-5.2,0.8);
    \draw[ thick,arrows = {-Stealth[fill=black,length=8pt]}](-5.8,-0.6)--(-5.9,-0.4);

    \coordinate(A1) at (-4.5,0);
    \coordinate(A2) at (-4.3,-0.6);
    \draw[ thick] (-5.5,-0.8) arc (-30:-60:-2);
    %\draw[xshift=5cm,white,double=blue,ultra thick,double distance=1.6pt] (-4.7,0) .. controls +(1,0) and +(-1,0) .. (-5,0);
\end{tikzpicture}
\vspace{1.5ex}
\end{subfigure}
\caption{\label{fig:oneloop-nonplanar}
One-loop amplitudes with two external legs: planar vs. non-planar Feynman diagrams.}
\end{figure}

Let us make an important remark. 
In noncommutative quantum field theory, at the one-loop level, one should consider a  supplementary non-planar Feynman diagram (i.e. with lines crossing over each other), on top of the standard one-loop diagram, 
as illustrated on the second diagram on fig.\ref{fig:oneloop-nonplanar}).

Such diagrams are actually at the origin of UV/IR mixing in NCQFTs and do not have  any classical equivalent \cite{Hersent:2020lsr} 
It has nevertheless been shown that the case of the non-commutative $\mathbb{R}_\kappa^{3}$ space-time has some simplifying propreties. Indeed, due to the braiding of the NCQFT as shown in \cite{Freidel:2005bb,Sasai:2007me}, such non-planar diagrams are equivalent to the standard planar one-loop diagram \cite{Sasai:2009jm}.  It is therefore enough to study the cutting rule for this planar one-loop Feynman diagram.
%Its integral form and after expanding it on characters.
%
%

%%%
\subsection{Feynman and Hadamard propagators on {$\R^{3}_{\ka}$}}
%\subsection{Deformed Euclidean Feynman and Hadamard propagators}
\label{sec:prop}
%%%

Let us start with the noncommutative field theory based on the $\SO(3)$ momentum space. The $\SU(2)$ case will be dealt with afterwards, and can be found at the end of this section.

Massive particles are encoded in 3D gravity as topological defects  parameterized by an angle $\varphi = \frac{m}{\kappa} \in [0,\pi/2]$, leading a renormalized mass $M = \kappa \sin\varphi$ \cite{Deser:1983tn,Matschull:1997du}.  This is this renormalized mass that enters the effective noncommutative field theory action \eqref{actionG}. The Feynman propagator can be read directly from the kinetic term of this action. In momentum space, it is given by \cite{Freidel:2005bb}:
\be
K_\varphi^{\rE}(g)
=
\f1{\vec{P}^{2}(g) - M^{2} + i \epsilon }
=
\f1{\kappa^{2}(\sin^{2}\theta - \sin^{2}\varphi + i \epsilon)}
\,.
\ee
Its Fourier transform gives the Feynman propagator in coordinate space,
\begin{equation}
    \cG_\varphi(\vec{x}) = \frac{1}{(2 \pi)^{3}} \int \frac{d^{3} \vec{P}}{\sqrt{1 - \frac{p^{2}}{\kappa^{2}}}}  \frac{e^{i \vec{x} \vec{P}}}{\vec{P}^{2} - M^{2}} = \frac{i}{4 \pi^{2} |\vec{x}|} \int^\kappa_{- \kappa} \frac{p dp}{\sqrt{1 - \frac{p^{2}}{\kappa^{2}}}} \frac{e^{-i |\vec{x}| p}}{p^{2} - M^{2}}
    \,.
\end{equation}
This propagator actually satisfies the Klein-Gordon equation with mass M and a $\delta_{0}(x)$-distribution source, which is the noncommutative counterpart of the standard $\delta$-distribution. It is the most localized distribution in coordinate space in the noncommutative $\R^{3}_{\ka}$,
\begin{equation}
    \delta_0(\vec{x}) = \frac{\kappa^{3}}{8 \pi } \int dg e^{i \vec{x} \vec{P}(g)} = \frac{1}{(2 \pi)^{3}} \int \frac{d^{3} \vec{P}}{\sqrt{1 - \frac{p^{2}}{\kappa^{2}}}} e^{i \vec{x} \vec{P}} = \frac{i \kappa^{3}}{4 \pi} \frac{J_1(\kappa |\vec{x}|)}{\kappa |\vec{x}|}
    \,,
\end{equation}
normalized to $\int \delta_0 = 1$, where $J_{1}$ is the standard Bessel function of the first kind.
Using this insight, it is possible to write the Feynman propagator as a convolution product in the radial coordinate  $r = |\vec{x}|$,
%
%Passing to radial coordinate $r = |\vec{x}|$, the propagator satisfies the radial Klein-Gordon equation with the source term given by $J_1$ . Therefore $r \cG_\varphi$ can be expressed as a convolution product:
\beq
\label{FeynpropxSO3}
    \cG_\varphi(r)
    &=& \frac{\kappa^{2}}{i 8 \pi M r} \int_{\R} d R \; e^{- i M|R|} J_1(\kappa(r - R)) \\
    & =& \frac{1}{4 \pi r \cos\varphi } e^{- iMr} - \frac{\kappa}{4 \pi r} \int_0^{\infty} dR \; \cos(M r) J_0(\kappa(r + R))\,.\nn
\eeq
%By getting rid of the absolute value and expanding the integral, we obtain the following: 
%\begin{equation}
%    \cG_\varphi(r) = \frac{1}{4 \pi r \cos(\varphi) } e^{- iMr} - \frac{\kappa}{4 \pi r} \int_0^{\infty} dR \; \cos(M r) J_0(\kappa(r + R))
%\end{equation}
The integral on the right vanishes as $r$ goes to infinity, and the leading order of the deformed Feynman propagator at large distances is given by the standard Feynman propagator, up to a $\cos\varphi$ numerical factor:
\be
    \cG_\varphi(r) \,\,\underset{r\rightarrow \infty}\sim\,\, \frac{e^{- i \kappa r \sin\varphi }}{4 \varphi r \cos\varphi}\,.
\ee
The integral term in the expression of the NCQFT propagator  thus describes the noncommutative correction to the standard QFT propagator. Note that it only affects its real part.

On the other hand,  at a short distances, when $r \to 0$, the behavior of the propagator is strongly affected by the space-time noncommutativity, which fully regularizes it. Indeed, the deformed Feynman propagator is finite at $r = 0 $ unlike its standard counterpart: 
\begin{equation}
\cG_\varphi(r)
\,\,\underset{r\rightarrow 0}\rightarrow\,\,
\frac{\kappa e^{- i \varphi}}{4 \pi \cos\varphi}
\,.
\end{equation}

\medskip

Another insightful expression of the Feynman propagator is given by its character expansion. Indeed, in momentum space, it is actually a (central) function over $\SO(3)$ and can thus be expanded over the characters of its irreducible representation, that is, in short, as a series over spins.
More precisely, an orthonormal basis of square-integrable functions on the Lie group $\SO(3)$, invariant under conjugation, is given by the characters $\chi_j$ for integer spins $j \in \mathbb{N}$. The character $\chi_j$ is the trace of the group elements in the spin-$j$ representation,
%They can be compute using Kirilov orbit method \cite{kirillov2004lectures}:
\begin{equation}
    \chi_j(g(\theta, \hat{u}) ) = \frac{\sin d_j \theta}{\sin\theta}
    \,,
\end{equation}
where $d_j= 2j +1$ is the dimension of the representation of spin $j$. We  expand the Feynman  propagator in this basis:
\begin{equation}
    K^{E}_\varphi(g) = \sum_{j \in \mathbb{N}} \cK^{E}_j(\varphi) \chi_{j}(g)
    \,,
\end{equation}
where the coefficient can be computed as a contour integral\footnotemark{},
\be
\cK^{E}_j(\varphi) = \int dg \; \chi_j(g) K^{E}_\varphi(g) =  \frac{2 e^{-i d_j(\varphi - i \epsilon)}}{\kappa^{2} \cos\varphi} 
\,.
\ee
\footnotetext{
We start from the explicit expressions of the Feynman propagator and characters as central functions on $\SO(3)$,
$\kappa^{2}\,K_\varphi^{\rE}(g)= 1+(\sin^{2}\theta- \sin^{2}\varphi + i \epsilon)$ and $\chi_{j}=\sin d_{j}\theta/\sin\theta$. 
Then, since the Haar measure on $\SO(3)$ for central functions can be simplified to $\rd g =\f1\pi\sin^{2}\theta\rd\theta$ for $\theta$ running from 0 to $2\pi$, we write the coefficients  of the spin expansion of the Feynman propagator as contour integrals in the complex plane,
\begin{equation*}
    \cK^{E}_j(\varphi) = \frac{1}{\kappa^{2} i \pi} \int_0^{2 \pi} \sin^{2}\theta \rd \theta \; \frac{e^{i (2j+1) \theta}}{\sin\theta}\frac{1}{ \sin^{2}\theta - \sin^{2}\varphi + i \epsilon} = \frac{-1}{2 \pi} \int_{\cC} dz \; z^{2j} \frac{X}{(X^{2} - \sin^{2}\varphi + i \epsilon)}
\end{equation*}
where $X = \sin(\theta) = \frac{z - z^{-1}}{2i}$ and $\cC$ is the unit cercle in $\mathbb{C}$.
The denominator can be expanded and we find poles around $\pm \exp{\pm i \varphi}$:
\begin{equation*}
    \frac{1}{X^{2} - \sin^{2}\varphi + i \epsilon} = \frac{(2iz)^{2}}{(z - e^{i \varphi} - (1 + i) \epsilon)(z - e^{- i \varphi} - (1 - i) \epsilon)(z - e^{-i \varphi} + (1 - i) \epsilon)(z + e^{i \varphi} + (1 + i) \epsilon)} 
\end{equation*}
We can then evaluate the integral using the residue formula. Since $\epsilon > 0$, only the poles in $\pm \exp{- i \varphi}$ contribute and we obtain the announced decomposition.
}
These coefficients $\cK^{E}_j(\varphi)$ are the quantized equivalent of the coordinate expression of the Feynman propagator. Indeed, as explained for example in \cite{Freidel:2005ec}, the spin $j$ is best interpreted as the quantized radial coordinate in the context of 3d quantum gravity and noncommutative coordinate operators. Comparing to the previous expression \eqref{FeynpropxSO3} for $\cG_\varphi(r)$, we indeed see that the expression for  $\cK^{E}_j(\varphi)$ is simpler and does not involve any extra term. So, in some sense, one has traded the integral correction in $\cG_\varphi(r)$ for a discrete length spectrum in $\cK^{E}_j(\varphi)$, both reflecting the effect of noncommutativity.

\medskip

Following the usual treatment of propagators in QFT, we define the Hadamard propagator from the imaginary part of the Feynman propagator:
\begin{equation}
\cH^{E}_\varphi(g)
\equiv
i K^{E}_\varphi(g) - i K^{E}_{- \varphi}(g)
=
\frac{4 \tan{\varphi}}{\kappa^{2}} \sum_{j\in\N} \chi_j(\varphi) \chi_j(g)
=
\frac{4 \tan{\varphi}}{\kappa^{2}}
\,
\tdelta_\varphi(g)
\,,
\end{equation}
where we recognize the Dirac distribution $\tdelta_\varphi$ on $\SO(3)$ fixing the class angle of the group element $g$ to the mass angle $\vphi$,
\be
\tdelta_\varphi(g)
=
\sum_{j\in\N} \chi_j(\varphi) \chi_j(g)
=
\frac12\Big{[}
\delta_\varphi(g) + \delta_{\pi- \varphi}(g)
\Big{]}
\,,
\ee
where the distribution $\delta_{\vphi}$ is the Dirac distribution fixing the class angle in $\SU(2)$:
\be
\delta_{\vphi}(g)
=
\int_{\SU(2)}\rd h\,\delta(gh\,g_{(\vphi,\hat{z})}^{-1}\,h^{-1})
=
\sum_{j\in\f\N2} \chi_j(\varphi) \chi_j(\theta)
\,,
\ee
As expected, the Hadamard propagator encodes the mass-shell condition on $\SO(3)$, which is the noncommutative equivalent of $\delta(p^{2} - m^{2})$ in standard QFT.
Let us underline the appropriate identification between the class angles $\vphi$ and $(\pi-\vphi)$, reflecting the identification of group elements $g\sim -g$. This identification restrict the sum from half-integers to integers, as expected.
More details on distributions on $\SO(3)$ versus $\SU(2)$ can be found in appendix \ref{appendixB}.

\medskip

We are now ready to tackle the case of a $\SU(2)$ momentum space and the corresponding $\SU(2)$  Feynman propagator. 
The main obstacle of working on $\SU(2)$ is that the momentum map $g\in\SU(2)\mapsto \vP(g)\in\R^{3}$ is not bijective but two-to-one, and thus there is no proper Fourier transform between $\cC(\mathbb{R}_\kappa^{3})$ and $\cC(\SU(2))$. Working on $\SO(3)$ actually allows to bypass this issue, by identifying the group elements $g\sim -g$. From this point of view, working on $\SU(2)$ instead of $\SO(3)$  means to be able to distinguish $g$ from $-g$. Some solutions to define an appropriate group Fourier transform have been proposed, e.g. lifting the 3d integrals to 4d in \cite{Freidel:2005ec}, or introducing an extra measure factor in \cite{Livine:2008sw}, or using the Schwinger parametrization in terms of spinorial variables \cite{Dupuis_2012}.
As we are interested here in the QFT propagators, the most straightforward method to lift all decomposition over integer spins to series over all half-integer spins.

Then, simply extending the previous series to half-integers, one can guess an appropriate Feynamn propagator on the $\SU(2)$ momentum space:
\begin{equation}
\label{offshellP1}
\sk^{E}_\varphi(g)
=
\frac{2}{\kappa^{2}} \sum_{j \in \mathbb{N}/2} e^{- i d_j (\varphi - i \epsilon)} \chi_j(\theta)
=
\frac{1}{2 \kappa^{2} \left(\sin^{2}\f\theta 2 - \sin^{2}\f\varphi 2 + i \epsilon\right)}
=
\frac{-1}{\kappa^{2}(\cos\theta - \cos\varphi) }
\,. 
\end{equation} 
Taking its imaginary part gives the corresponding Hadamard propagator:
\begin{equation}
\sh^{E}(g)
\equiv
i k_\varphi^{E}(g) - i k_{-\varphi}^{E}(g)
= \frac{4 \sin\varphi}{\kappa^{2}} \delta_\varphi(g)
   \,,
\end{equation}
where we recover the expected mass-shell condition, fixing the class angle of the group element $g$ in $\SU(2)$.
As wanted, we have lifted the identification the $\SO(3)$ identification $g\sim -g$, or equivalently $\vphi\sim \pi-\vphi$. The new $\SU(2)$ Feynman propagator amounts to modifying the kinetic term of our original NCQFT,
\be
S[\phi]
=
\frac{1}{2} \int_G \rd g \; \cK(g) \phi(g) \phi(g^{-1}) + \frac{\lambda}{3} \int_G [\rd g]^{3} \; \delta(g_1 g_2 g_3) \phi(g_1)\phi(g_2)\phi(g_3) 
\ee
from the original theory written in \cite{Freidel:2005bb,Freidel:2005me},
\be
\cK_{\SO(3)}(g)
=P^{2}(g)-M^{2}
=\ka^{2}(\sin^{2}\theta-\sin^{2}\vphi)
=-\f{\ka^{2}}4\left[\chi_{1}(g)-\chi_{1}(\vphi)\right]
\ee
to a new version based on the character of spin $\f12$ instead,
\be
\cK_{\SU(2)}(g)
=-\ka^{2}\left[\chi_{\f12}(g)-\chi_{\f12}(\vphi)\right]
=-2\ka^{2}(\cos\theta-\cos\vphi)
\,.
\ee
This $\SU(2)$ propagator version was actually derived in previous work from 3d quantum gravity in the group field theory formalism \cite{Fairbairn_2007} and interpreted as filling the discrete Ponzano-Regge path integral with spin-$\f12$ defects \cite{Livine:2011yb,BenGeloun:2018eoe}.

However we will go further than this propagator ansatz and add a extra term. We will see later in section \ref{loopESU2} that this new term is necessary to ensure unitarity at one-loop. Explicitly, we introduce an enhanced Feynman propagator:
\begin{equation}
\label{offshellP2}
\kEphi(g)
=
\frac{2}{\kappa^{2}} \sum_{d_j \in \mathbb{N}} e^{- i d_j (\varphi - i \epsilon)} \chi_j(\theta)
=
\sk^{E}_{\vphi}(g) + \frac{1}{\kappa^{2}\sin \theta}
%=
%\frac{1}{2 \kappa^{2} (\sin^{2}(\theta /2) - \sin^{2}(\varphi/2) + i \epsilon} + \frac{1}{\kappa^{2}\sin \theta}
\,,
\end{equation} 
where we have added a term interpretable as the $\SU(2)$-representation with vanishing $d_{j}=0$. Actually, it is an infinite-dimensional representation with Casimir $\vec{J}^{2}=-\f14$, thus corresponding to a negative spin $j=-\f12$. Its character would be $\chi_{-\f12}(\theta) = \frac{1}{2 \sin \theta}$ and diverges at the identity $g=\id$, $\theta=0$.
Nevertheless, this divergence goes away with the Haar measure and  only leads to a constant offset {of $2 / \pi$} of the Feynman propagator after a Fourier transform. 
This new 0-mode term does not affect the relation  between the Feynman and Hadamard propagators,
\begin{equation}
   i {\mathbf k}_\varphi^{E}(g) - i {\mathbf k}_{-\varphi}^{E}(g) = \sh^E_{\vphi}(g)
   \,,
\end{equation}
so it is a legimitate ansatz for a Feynman propagator.
Finally, in the commutative limit where the deformation parameter is sent to infinity, $\kappa \to \infty$, this extra term vanishes, $(\kappa^{2} \sin \theta )^{-1} \to 0$ and we recover as wanted the standard Feynman propagator of the undeformed QFT. Nevertheless, the physical interpretation of this extra 0-mode is not quite clear, it translates, after inversion, into a more complicated kinetic term in the field theory action, and we will discuss its role more in section \ref{sec:dicussion}.

\subsection{One-loop unitary: $\SO(3)$ Feyman propagator}
%\subsection{One-loop correction and Euclidean unitarity}
%%%

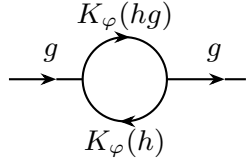
\begin{figure}[htb!]
\begin{tikzpicture}[scale=0.7]
    \coordinate(O) at (-5.3,0);
    \coordinate(5) at (-7.5,0);
    \coordinate(6) at (-6.1,0);
    \coordinate(7) at (-6.6,0);
    \coordinate(52) at (-4.5,0); %11.3
    \coordinate(62) at (-3.4,0);
    \node[scale=1.2] at (-6.7,0.5) { $g$};
    \node[scale=1.2] at (-3.6,0.5) { $g$};
    
    \draw[ thick,arrows = {-Stealth[fill=black,length=6pt]}](5)--(7);
    \draw[ thick](7)--(6);
    
    \draw[ thick,arrows = {-Stealth[fill=black,length=6pt]}](52)--(62);
    \coordinate(72) at (-3,0);
    \draw[ thick](62)--(72);
    
	\draw[ thick] (O) circle (0.8) ;

    \node[scale=1.2] at (-5.3,1.2) { $K_\varphi(hg)$};
    \node[scale=1.2] at (-5.3,-1.2) { $K_\varphi(h)$};

    \draw[ thick,arrows = {-Stealth[fill=black,length=6pt]}](-5.3,0.8)--(-5.2,0.8);
    \draw[ thick,arrows = {-Stealth[fill=black,length=6pt]}](-5.3,-0.8)--(-5.4,-0.8);
\end{tikzpicture}
\caption{\label{fig:oneloopE}
The one-loop diagram, with incoming group momentum  $g$. The corresponding vector momentum is $\vP(g)$ with norm $|\vP|= \kappa \sin\theta$. The momentum circulating around the loop is $h$, while the mass of the non-commutative scalar field is $M = \kappa \sin\varphi$.
In the commutative limit, sending the Planck mass to infinity, $\kappa \to \infty$, one keeps the mass $M$ fixed and finite while sending accordingly  the mass angle $\varphi$  to $0$.
The deformed Feynman diagram amplitudes then give back the standard QFT integrals.
}
\end{figure}

Now that the propagators are well-defined for this Euclidean field theory, let us move on to
the computation of the one-loop integrals as illustrated on figure \ref{fig:oneloopE} with on-shell and off-shell propagators. This will allow us to check the validity (or not) of the Cutkosky's cut rules for the non-commutative field theory with $\SO(3)$ momenta.

\begin{figure}[htb!]
\centering
\begin{subfigure}[b]{0.4\textwidth}
\centering    \includegraphics[height=40mm]{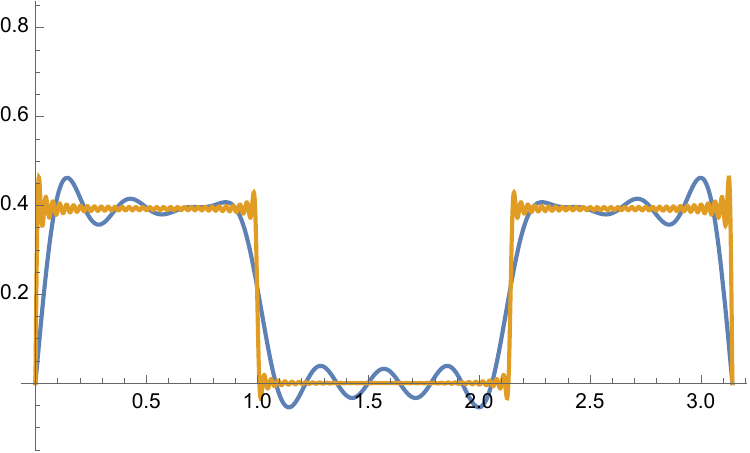}
\caption{Mass angle $\vphi=0.5<\pi/4$.}
\end{subfigure}
\hspace*{5mm}
\begin{subfigure}[b]{0.4\textwidth}
\centering    \includegraphics[height=40mm]{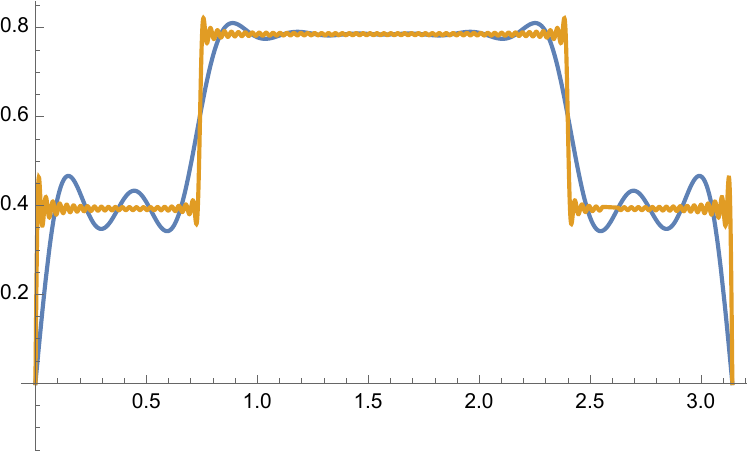}
\caption{Mass angle $\vphi=1.2>\pi/4$.}
\end{subfigure}
\caption{\label{fig:plotSO3}
Plot for $\theta\in[0,\pi]$ of the sum $\sum_{j }^{J} (d_j)^{-1} \sin\theta \sin^{2}d_{j}\vphi$ over integer spins $j\in\N$, with two cut-offs $J=10$ (in blue) and $J=100$ (in orange) for values of mass angles $\vphi$ below $\pi/4$ (on the left hand side) and above $\pi/4$ (on the right hand side). One clearly sees  the thresholds at $\theta=2\vphi$ and $\theta=\pi-2\vphi$ (respecting the $\Z_{2}$ symmetry of $\SO(3)$), and the three step levels with values 0, $\pi/8\sim0.39$ and $2\pi/8$.
}
\end{figure}

Let us fix the mass angle $\vphi\in[0,\pi/2]$. Indeed masses $\vphi$ and $\pi-\vphi$ are equivalent so we can safely restrict from the interval $[0,\pi]$ to $[0,\pi/2]$.
Let us then start by analyzing the one-loop integral with on-shell propagators. Its expansion on characters reads:
\begin{equation}
    \cL^{0,E}_\varphi(g)
    =
    \frac{i \kappa^{3}}{16 \pi}  \int \rd h \; \cH^{E}_\varphi(h)\cH^{E}_\varphi(hg)
    = \frac{i \tan^{2}\varphi}{\pi\kappa} \sum_{j \in \mathbb{N}} \frac{1}{d_j}  \chi_j(\theta)\chi_j^{2}(\varphi)
=
\frac{i}{\pi\kappa\cos^{2}\varphi \sin\theta} \sum_{j \in \mathbb{N}} \frac{1}{d_j} \sin d_{j}\theta \sin^{2}d_{j}\vphi
    \,.
\end{equation}
This expression  can  be explicitly computed either from the sum over spins or the integral over the group. Details are given  in appendix \ref{appendixB}.  We must distinguish different cases according to the value of the mass angle $\vphi$, as shown on the plots in figure \ref{fig:plotSO3}.

For a mass angle $\psi \in [0, \tfrac\pi4]$, we get:
%The result reads, for a momentum $\theta \in [0, \pi/2]$ and ~:
\begin{equation} 
    \mathcal{L}_\varphi^{0}(g) = \left\{
\begin{array}{cl}
\displaystyle\frac{i}{8 \kappa \cos^{2}\varphi \sin\theta}  &
\qquad \text{if} \quad 0 <\theta < 2 \varphi 
\,,\vspace*{1mm}\\
        0  & \qquad \text{if} \quad 2 \varphi< \theta <\displaystyle\f\pi2
        \,,
    \end{array}
\right.
\end{equation}
while, for  a mass angle $\psi \in [ \tfrac\pi4,\tfrac\pi2]$, we get a extra contribution:
\begin{equation} 
    \mathcal{L}_\varphi^{0}(g) = \left\{
\begin{array}{cl}
\displaystyle\frac{i}{8 \kappa \cos^{2}\varphi \sin\theta}  &
\qquad \text{if} \quad 0 < \theta < \pi - 2 \varphi 
\,,\vspace*{1mm}\\
         \displaystyle \frac{i}{4 \kappa \cos^{2}\varphi \sin\theta} 
 & \qquad \text{if} \quad  \pi - 2 \varphi< \displaystyle\f\pi2
        \,.
    \end{array}
\right.
\end{equation}
Let us keep in mind that these expressions are invariant under $\theta \leftrightarrow \pi-\theta$.
Comparing with standard undeformed field theory, the sector $0 < \theta < 2 \varphi$  corresponds to the sector $|p| < 2 m $, while there is no classical  counterpart to the ``IR-UV'' identification $\theta \leftrightarrow \pi-\theta$ due to working with $\SO(3)$ momenta.

We now compute  the off-shell one-loop with incoming momentum $g$, internal momentum $h$ and mass angle $\vphi$, as illustrated on fig.\ref{fig:oneloopE}.

Using the character expansion of the Feynman propagator, this one-loop integral with off-shell propagators read:
\beq
\cL^{E}_\varphi(g)
&=&
\frac{i \kappa^{3}}{8 \pi} \int dh \; K^{E}_\varphi(h) K_\varphi^{E}(hg)
=
\frac{i \kappa^{3}}{8 \pi}
\sum_{j,j' \in \mathbb{N}} 
\cK^{E}_j(\varphi)\cK^{E}_{j'}(\varphi)
\int dh \;  \chi_{j}(h) \chi_{j'}(hg)
\nn\\
&=&
\frac{i \kappa^{3}}{8 \pi}
\sum_{j \in \mathbb{N}} 
\f{1}{d_{j}}\left(\cK^{E}_j(\varphi)\right)^{2}
\chi_{j}(g)
=
\frac{i}{2 \pi \kappa \cos^{2}\varphi\sin\theta} \sum_{j \in \mathbb{N}} \frac{e^{- i d_j 2 (\varphi - i \epsilon)}}{d_j} \sin d_{j}\theta
%\frac{i}{2 \pi \kappa \cos^{2}(\varphi)} \sum_{j \in \mathbb{N}} \frac{e^{- i d_j 2 (\varphi - i \epsilon)}}{d_j} \chi_j(\theta) 
\,.
\label{eq:LEg}
\eeq
%\begin{equation}
%    \cL^{E}_\varphi(g) = \frac{i \kappa^{3}}{8 \pi} \int dh \; K^{E}_\varphi(h) K_\varphi^{E}(hg) = \frac{i}{2 \pi \kappa \cos^{2}(\varphi)} \sum_{j \in \mathbb{N}} \frac{e^{- i d_j 2 (\varphi - i \epsilon)}}{d_j} \chi_j(\theta) 
%\end{equation}
%
%
\begin{figure}[htb!]
\centering
\begin{subfigure}[b]{0.4\textwidth}
\centering    \includegraphics[height=40mm]{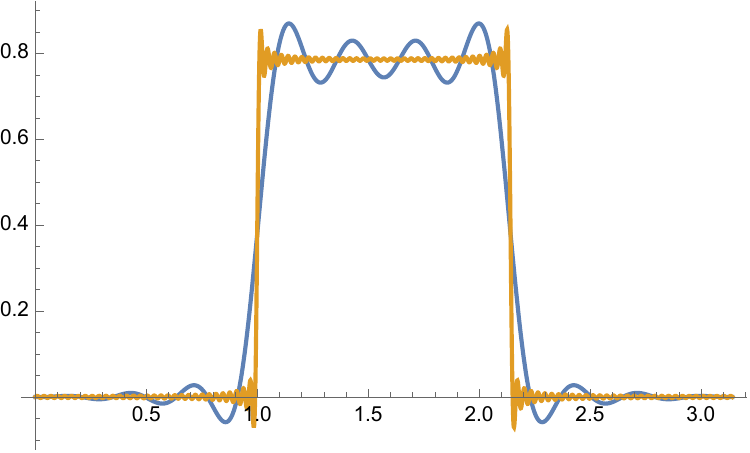}
\caption{Mass angle $\vphi=0.5<\pi/4$.}
\end{subfigure}
\hspace*{5mm}
\begin{subfigure}[b]{0.4\textwidth}
\centering    \includegraphics[height=40mm]{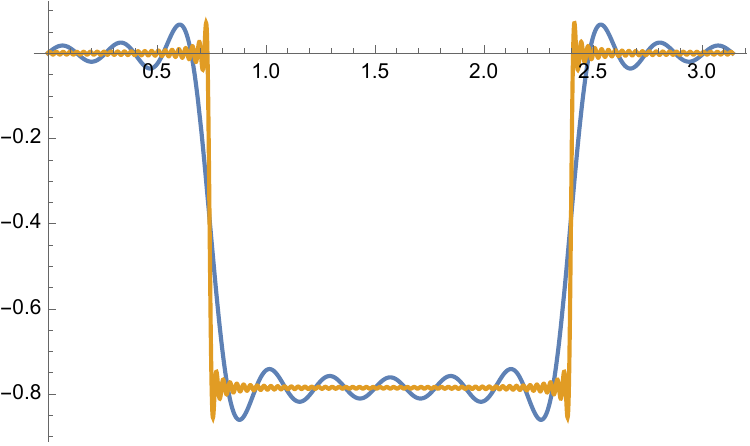}
\caption{Mass angle $\vphi=1.2>\pi/4$.}
\end{subfigure}
\caption{\label{fig:plotSO3-FeynR}
Plot for $\theta\in[0,\pi]$ of the real part $\sum_{j }^{J} (d_j)^{-1}e^{-2d_{j}\eps}\,  \cos 2d_{j}\vphi\,\sin\theta$  of the series giving the one-loop Feynman diagram $\cL^{E}(g)$ in the formula \eqref{eq:LEg}. We sum up to integer spins $j\in\N$, with two cut-off values $J=10$ (in blue) and $J=100$ (in orange) for values of mass angles $\vphi$ below $\pi/4$ (on the left hand side) and above $\pi/4$ (on the right hand side). One clearly sees  the thresholds at $\theta=2\vphi$ and $\theta=\pi-2\vphi$ (respecting the $\Z_{2}$ symmetry of $\SO(3)$), and the flip of sign once the mass angle $\vphi$ grows larger than $\pi/4$.
}
\end{figure}
\begin{figure}[htb!]
\centering
\begin{subfigure}[b]{0.4\textwidth}
\centering    \includegraphics[height=40mm]{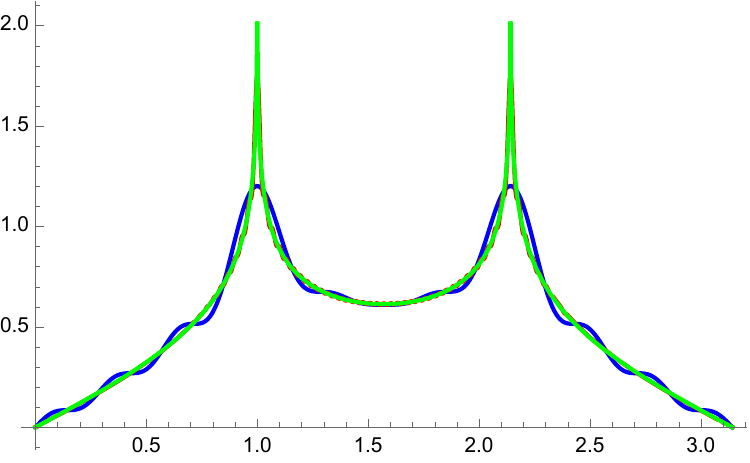}
\caption{Mass angle $\vphi=0.5<\pi/4$.}
\end{subfigure}
\hspace*{5mm}
\begin{subfigure}[b]{0.4\textwidth}
\centering    \includegraphics[height=40mm]{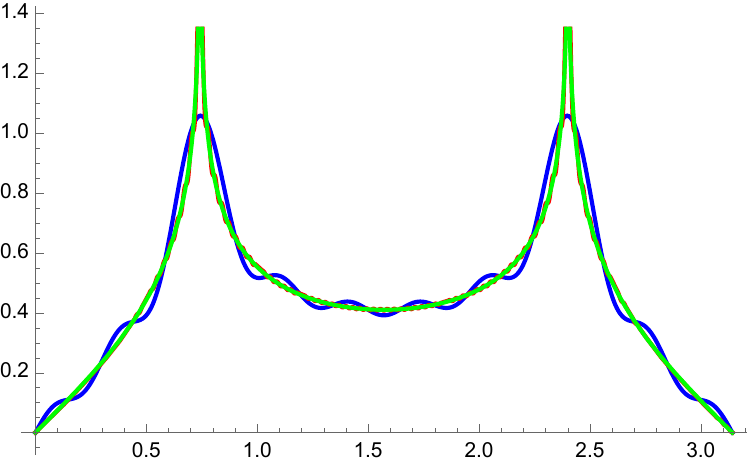}
\caption{Mass angle $\vphi=1.2>\pi/4$.}
\end{subfigure}
\caption{\label{fig:plotSO3-FeynI}
Plot for $\theta\in[0,\pi]$ of the imaginary part $\sum_{j }^{J} (d_j)^{-1}e^{-2d_{j}\eps}\,  \sin 2d_{j}\vphi\,\sin\theta$  of the series giving the one-loop Feynman diagram $\cL^{E}(g)$ in the formula \eqref{eq:LEg}. We sum up to integer spins $j\in\N$, with two cut-off values $J=10$ (in blue) and $J=100$ (in red), and plot the limit curve $\ln|(\sin2\varphi - \sin\theta)/(\sin2\varphi + \sin\theta)|$ (in green).
The curve in red is slightly hidden by the green curve, showing an excellent convergence of the series as the cut-off grows.
We give an example of a mass angle $\vphi$ below $\pi/4$ on the left hand side and above $\pi/4$ on the right hand side.
}
\end{figure}

We notice the same pre-factor in $i(\pi \kappa \cos^{2}\varphi\sin\theta)^{-1}$ as for the on-shell integral. We plot the real and imaginary parts of the series in figures \ref{fig:plotSO3-FeynR} and \ref{fig:plotSO3-FeynI}, in order to visualize the various cases. The series can be summed exactly and yields
for $\varphi\in[0, \pi/4]$:
\begin{equation}
\label{equ:offshellso(3)} 
\cL^{E}_\varphi(g) = \left\{
    \begin{array}{ll}
\displaystyle\frac{-1}{ 8 \kappa \cos^{2}\varphi \sin\theta}
\log\displaystyle\left|\frac{\displaystyle \sin2\varphi - \sin\theta}{\displaystyle\sin2\varphi + \sin\theta}\right|\,,
&\quad \textrm{if} \quad 0 < \theta < 2 \varphi 
\,,\vspace*{2mm}\\
\displaystyle\frac{-1}{ 8 \kappa \cos^{2}\varphi \sin\theta}
\log\displaystyle\left|\frac{\displaystyle \sin2\varphi - \sin\theta}{\displaystyle\sin2\varphi + \sin\theta}\right|
+ \frac{i}{8\kappa \cos\varphi^{2} \sin\theta}
\,,
&\quad\textrm{if} \quad  2 \varphi< \theta < \displaystyle\f\pi2
\,.
\end{array}
\right.
\end{equation}
For a mass angle in the upper half-range  $\varphi\in[\pi/4,\pi/2]$, we get:
\begin{equation}
\cL^{E}_\varphi(g) = \left\{
    \begin{array}{ll}
\displaystyle\frac{-1}{ 8 \kappa \cos^{2}\varphi \sin\theta}
\log\displaystyle\left|\frac{\displaystyle \sin2\varphi - \sin\theta}{\displaystyle\sin2\varphi + \sin\theta}\right|\,,
&\quad \textrm{if} \quad 0 < \theta < \pi-2 \varphi 
\,,\vspace*{2mm}\\
\displaystyle\frac{-1}{ 8 \kappa \cos^{2}\varphi \sin\theta}
\log\displaystyle\left|\frac{\displaystyle \sin2\varphi - \sin\theta}{\displaystyle\sin2\varphi + \sin\theta}\right|
- \frac{i}{8\kappa \cos\varphi^{2} \sin\theta}
\,,
&\quad\textrm{if} \quad  \pi-2 \varphi< \theta < \displaystyle\f\pi2
\,,
\end{array}
\right.
\end{equation}
where the only notable change is the switch of sign in the imaginary part for higher values of the incoming momentum $\theta$. This is simply due to a sign switch of the expression in the logarithm.

At the end of the day, comparing the off-shell and on-shell integrals, we see that the Euclidean version of the Cutkosky cut rule is valid in the $\SO(3)$ noncommutative field theory at small masses, with  $ \varphi \in [0, \pi/4]$ (corresponding to ${m < (16 G)^{-1}} = (8 m_p)^{-1} $ where $m_p$ is the Planck mass) 
\be
i\textrm{Im}\,\cL^{E}_\varphi(g) +\cL^{0}_\varphi(g) 
=
\frac{i}{8 \kappa \cos^{2}\varphi \sin\theta}\,,
\ee
with the imaginary part of the off-shell integral vanishing for momenta $\theta\in[0,2\varphi]$ and $\theta\in[\pi - 2 \varphi,\pi/2]$, while the on-shell integral vanishes for all other values of the momentum $\theta\in[2\varphi, \pi - 2 \varphi]$.
On the other hand, once the mass is large enough, for $2 \varphi \in [\pi/2, \pi]$, this cut rule is violated and unitarity fails, due to the new non-vanishing term in the on-shell amplitudes for momentum $\theta\in[2\varphi, \pi - 2 \varphi]$.
In the following section, we will see how one can fix this problem by properly working with a $\SU(2)$ momentum space instead of $\SO(3)$.

%To conclude this part, in SO(3), we get an Euclidean Cutkosky rule valid for small masses when  $ 2 \varphi \in [0, \pi/2]$  i.e., in comparison to Planck mass, particles must have a mass $m < (16 G)$. Apparition of the sector $\pi - 2 \varphi < \theta < \pi$ is due to the structure of SO(3) and the identification between element $g \longleftrightarrow -g $. It should disappear when moving to SU(2), the double cover of SO(3). At the same time, it seems to behave un-classically in the more massive sector (i.e., $2 \varphi \in [\pi/2, \pi]$). Since the angle $\varphi$ corresponds to a deficit angle of $2 \varphi$, this un-classical sector physically corresponds to a collision between two particles that meet and form a deficit angle of $2 \varphi$. 

%%%
\subsection{One-loop unitary: $\SU(2)$ Feyman propagator}
\label{loopESU2}
%\subsection{One-loop correction and Euclidean unitarity}
%%%

\begin{figure}[htb!]
\centering
\begin{subfigure}[b]{0.4\textwidth}
\centering    \includegraphics[height=30mm]{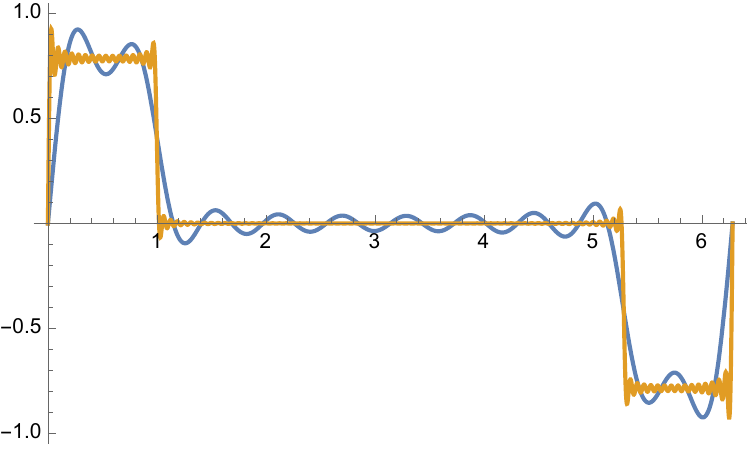}
\caption{Mass angle $\vphi=0.5<\pi/4$.}
\end{subfigure}
\hspace*{5mm}
\begin{subfigure}[b]{0.4\textwidth}
\centering    \includegraphics[height=30mm]{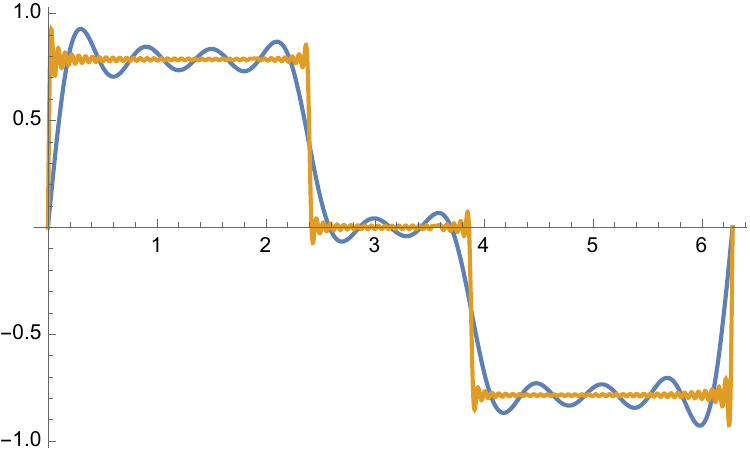}
\caption{Mass angle $\vphi=1.2>\pi/4$.}
\end{subfigure}
\caption{\label{fig:plotSU20}
Plot for $\theta\in[0,2\pi]$ of the sum $\sum_{j}^{J}d_j^{-1} \sin d_{j}\theta \sin^{2}d_{j}\vphi$  for the one-loop  diagram $\ell_\phi^{0,E}(g)$ with $\SU(2)$ Hadamard propagators. We sum up to integer spins $j\in\N$, with two cut-off values $d_{J}=10$ (in blue) and $d_{J}=100$ (in orange) for values of mass angles $\vphi$ below $\pi/4$ (on the left hand side) and above $\pi/4$ (on the right hand side). 
}
\end{figure}

Let us now look into the one-loop amplitudes based on the $\SU(2)$ momentum space. The on-shell diagram uses the Hadamard propagator, given by the distribution $\delta_{\vphi}(g)$ fixing the class angle in $\SU(2)$,
\be
\ell^{0,E}_\varphi(g)
=
\frac{i \kappa^{3}}{16 \pi}
\int \rd h \, \sh^E_\varphi(h) \sh^E_\varphi(hg)
=
\frac{i \sin^{2}\varphi}{\pi \kappa}
\sum_{j \in \f{\mathbb{N}}2}
\frac{1}{d_j} \chi_j^{2}(\varphi) \chi_j(\theta) 
=
\frac{i}{\pi\kappa\sin\theta}
\sum_{j \in \f{\mathbb{N}}2}
\frac{1}{d_j} \sin d_{j}\theta \sin^{2}d_{j}\vphi
\,,
\ee
where we underline that the main difference with the $\SO(3)$ case is that we now sum over all half-integer spins $j$.
This series can be summed exactly. This gives us for $\vphi\in[0,\f\pi2]$ and $\theta\in[0,\pi]$:
\be
    \ell_\phi^{0,E}(g) = \left\{
    \begin{array}{cl}
       \displaystyle\frac{i}{4 \kappa  \sin\theta}  & \qquad \text{if} \quad 0 < \theta < 2 \varphi\,,
       \vspace*{2mm}\\
        0  &\qquad \text{if} \quad 2 \varphi< \theta < \pi \,,
    \end{array}
\right.
\label{equ:SU(2)onshell}
\ee

where we see the expected threshold at $\theta_{c}=2\vphi$, as shown on figure.\ref{fig:plotSU20}, with a step drop from the constant value $\pi/4$ down to 0.

We can compute this series for higher angles $\theta\in[\pi,2\pi]$. This gives an inverted behavior with a second threshold at $2\pi-2\vphi$, as illustrated on figure \ref{fig:plotSU20}. Let us nevertheless point out that we can restrict ourselves to $\theta\in[0,\pi]$ to avoid redundacy in our angle-axis parametrization of $\SU(2)$ group elements, and thus discard higher values of the angle $\theta\in[\pi,2\pi]$ as physically irrelevant.

\begin{figure}[htb!]
\centering
\begin{subfigure}[b]{0.4\textwidth}
\centering    \includegraphics[height=30mm]{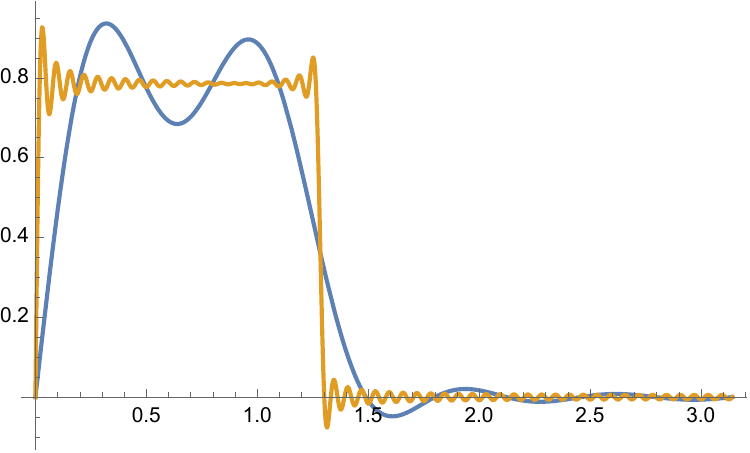}
%\caption{Mass angle $\vphi=0<\pi/4$.}
\end{subfigure}
\hspace*{5mm}
\begin{subfigure}[b]{0.4\textwidth}
\centering    \includegraphics[height=30mm]{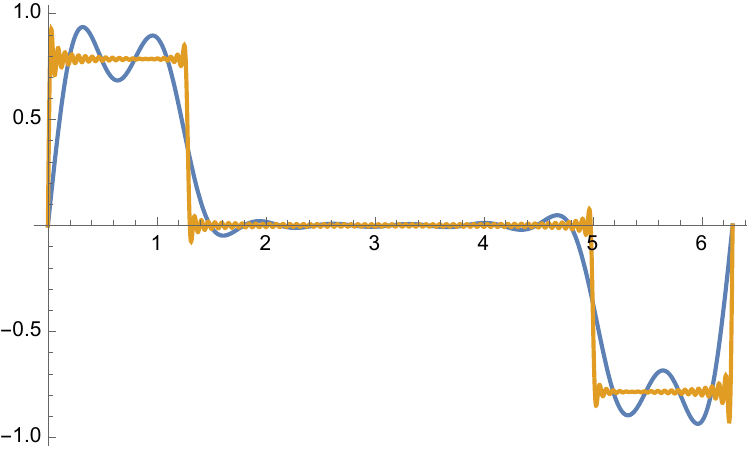}
%\caption{Mass angle $\vphi=1.2>\pi/4$.}
\end{subfigure}
\caption{\label{fig:plotSU20higher}
Plot of the sum $\sum_{j}^{J}d_j^{-1} \sin d_{j}\theta \sin^{2}d_{j}\vphi$ for $\vphi=2.5\in[\pi,2\pi]$, and cut-off values $d_{J}=10$ (in blue) and $d_{J}=100$ (in orange),
the left plot for a angle range $\theta\in[0,\pi]$ and the right plot the whole range $\theta\in[0,2\pi]$. We clearly see the thresholds at $2\pi-2\vphi$ and $2\vphi$.
}
\end{figure}

Finally, for higher mass angle  $\vphi\in[\f\pi2,\pi]$, the profile stays the same but the two thresholds $\theta_{c}$ at $2\vphi$ and $2\pi-2\vphi$ now plays inverse roles, as shown on figure \ref{fig:plotSU20higher}.

\medskip

\begin{figure}[htb!]
\centering
\begin{subfigure}[b]{0.5\textwidth}
\centering    \includegraphics[height=30mm]{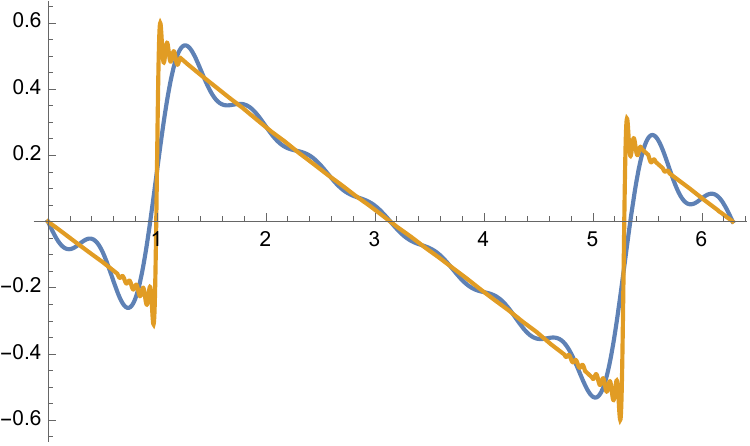}
\caption{The linear evolution with non-zero slope is typical of a non-trivial monodromy. This is the term supposed to satisfy the optical theorem with the on-shell amplitude.}
\end{subfigure}
\hspace*{5mm}
\begin{subfigure}[b]{0.4\textwidth}
\centering    \includegraphics[height=30mm]{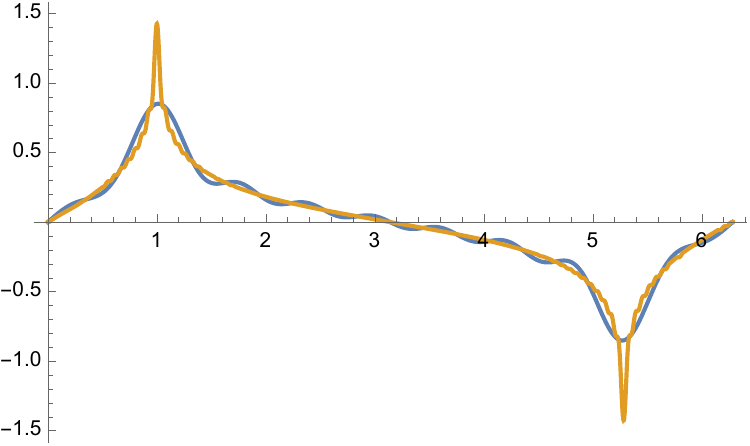}
\caption{This is the purely quantum contribution to the off-shell amplitude based on the diagram with Feynman propagators.}
\end{subfigure}

\caption{\label{fig:plotSU2Feynnaive}
Plot of the real and imaginary parts of the series defining the one-loop Feynman diagram amplitude \eqref{eq:lEg},  $\f12\sum_{j}^{J}d_j^{-1}e^{-2d_{j}\eps} \sin d_{j}\theta \cos 2d_{j}\vphi$ on the left and $\f12\sum_{j}^{J}d_j^{-1}e^{-2d_{j}\eps} \sin d_{j}\theta \sin 2d_{j}\vphi$ on the right,
 for a regulator $\eps=10^{-5}$, and cut-off values $d_{J}=10$ (in blue) and $d_{J}=100$ (in orange), and a mass angle $\vphi=0.5$.
}
\end{figure}

\begin{figure}[htb!]
\centering

\begin{subfigure}[b]{0.4\textwidth}
\centering    \includegraphics[height=30mm]{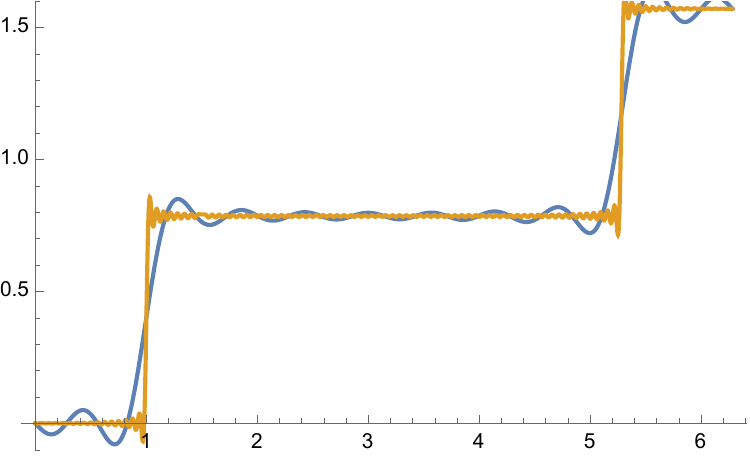}
%\caption{Mass angle $\vphi=1.2>\pi/4$.}
\end{subfigure}
\hspace*{5mm}
\begin{subfigure}[b]{0.4\textwidth}
\centering    \includegraphics[height=30mm]{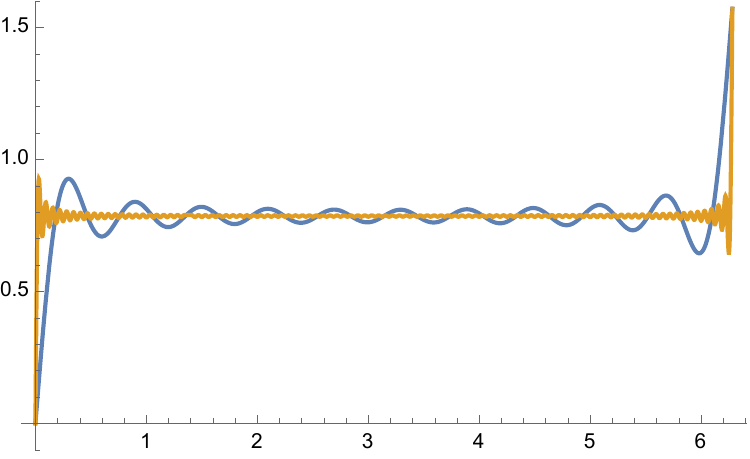}
%\caption{Mass angle $\vphi=1.2>\pi/4$.}
\end{subfigure}

\caption{\label{fig:plotSU2naive}
We correct the imaginary part of the Feynman diagram amplitude (corrected by a factor $\sin\theta$) with a linear term $\theta/4$  to remove its decreasing slope (on the left), and then add the on-shell amplitude to find a constant, thus yielding the Euclidean equivalent of the optical theorem. The linear term is the anomalous term forcing us to correct the Feynman propagator.}
\end{figure}

The computation of the off-shell Feynman diagram is more subtle due to the possible ambiguity in the definition of the Feynman propagator, as discussed earlier in section \ref{sec:prop}.
Indeed, it is possible to change the real part of the Feynman propagator while respecting the constraint that its imaginary part reamins the Hadamard propagator giving the mass-shell condition.
Let us first try the ``na\"ive'' ansatz \eqref{offshellP1}, given explicitly by
\be
k^{E}_\varphi(g)
=
\frac{2}{\kappa^{2}} \sum_{j \in \f\N2} e^{- i d_j (\varphi - i \epsilon)} \chi_j(\theta)
=
%\frac{-2}{\kappa^{2}(\cos\theta - \cos\varphi) }
\frac{1}{\kappa^{2} \left(\sin^{2}\f\theta 2 - \sin^{2}\f\varphi2 + i \epsilon\right)}
\,,
\nn
\ee
then the one-loop integral yields,
\be
\label{eq:lEg}
 \ell^{E}_\varphi(g)
 =
 \frac{i \kappa^{3}}{8 \pi} \int \rd h \; k_\varphi^{E}(h) k_\varphi^{E}(hg)
 =
 \frac{i}{2 \pi \kappa}\, \sum_{j \in \f\N2} \frac{e^{-i d_j 2 (\varphi - i \epsilon)}}{d_j} \chi_j(\theta)
 \,.
\ee
This series can be summed exactly\footnotemark. We plot its real and imagniary parts in figure \ref{fig:plotSU2Feynnaive}.
\footnotetext{
$\ell_\phi^{E}(g) $ can be computed by integrating $k_\varphi^{E}\propto 1/(\cos\theta - \cos\varphi)$ over $\varphi$. This gives a term in $\log[(1 - \cos)/(1+\cos)]$ which can be re-written as: 
\begin{equation}
    \log\left(\frac{\sin^{2}\varphi - \sin^{2}\f\theta2 + \sin^{2}(\varphi - \f\theta2)}{\sin^{2}\varphi - \sin^{2}\f\theta2 + \sin^{2}(\varphi + \f\theta2)}\right)
    =
    \log\left(1 - \frac{\sin2 \varphi \,\sin\theta}{\sin^{2}\varphi - \sin^{2}\f\theta2 + \sin^{2}(\varphi - \f\theta2)}\right)
    \nn\,.
\end{equation}
} 
However one find that it violates the Euclidean version of the optical theorem:
%and we get $\varphi \in [0, \pi/2)$ and varying from $0$ to $\pi$: 
%\begin{equation}
% \pi\ka \,  \text{Im}\,\ell^{E}_{\varphi}(p) +  \frac{\theta}{4 \sin\theta } -i\pi\ka \ell^{0, E}_\varphi(p) = \frac{\pi}{4 \sin\theta } 
%    \,,
%\end{equation}
\begin{equation}
   i\text{Im}\,\ell^{E}_{\varphi}(p) +  \ell^{0, E}_\varphi(p) = \frac{\pi}{4  \kappa \sin\theta }  -  \frac{i\theta}{4  \pi\kappa \sin\theta } 
    \,,
\end{equation}
with the expected $1/\sin\theta$ term on the right hand side, corresponding to the $1/|p|$ term in our Euclidean version of the cut rule in standard commutative QFT, but now with an extra anomalous term $\theta/\sin\theta$, as illustrated by the plots in figure \ref{fig:plotSU2naive}. 

\medskip

Making this theory unitary thus require to kill this anomaly.
As we announced in the previous section, we have identified an appropriate modification of the Feynman propagator that compensate this anomalous term, by introducing an extra 0-mode term as given in \eqref{offshellP2}, which we remind here:
\begin{equation}
\kEphi(g)
=
\frac{2}{\kappa^{2}} \sum_{d_j \in \mathbb{N}} e^{- i d_j (\varphi - i \epsilon)} \chi_j(\theta)
=
%\frac{1}{\kappa^{2} (\sin^{2}(\theta /2) - \sin^{2}(\varphi/2) + i \epsilon} + \frac{1}{\kappa^{2}\sin \theta}\,,
k^{E}_\varphi(g) + \frac{1}{\kappa^{2}\sin \theta}\,.
\nn
\end{equation} 
This new term has a pole at $\theta=0$, i.e. on the vanishing momentum $g=\id$. This creates a resonance at zero mass, thus we identify as an extra ``0-mode''.
%
%\begin{figure}[htb!]
%   \includegraphics[width=60mm]{NCQFT-Unitarity (Valentine)/offSU(2).pdf}
%   \caption{{\bf WHERE IS THIS FIG CITED IN TEXT? COULD YOU WRITE WHAT IS COMPUTED and WHAT ARE THE AXIS?}
%   \textcolor{red}{A representation of the of the real part of the SU(2) off-shell one loop with $\kappa = 1$.  Peaks indicate the changed of sign inside the logarithm of expression \eqref{offshellSU(2)} and therefore the apparition of the imaginary term. The orange part indicate a imaginary part null and the red a imaginary part non zero.} }
%\end{figure}
This modification of the the Feynman propagator compensates exactly the anomaly in the one-loop integral,
%
%We can compute exactly the resulting one-loop integral,
\be
\Bell^{E}_\varphi(g)
=
\frac{i \kappa^{3}}{8 \pi} \int dh \; \kEphi(h) \kEphi(hg)
=
\frac{i \kappa^{3}}{8 \pi} \int dh \; \sk_\varphi^{E}(h) \sk_\varphi^{E}(hg)
+ 
\frac{i \theta}{4 \pi \kappa \sin(\theta) }
=
\ell^{E}_\varphi(g)+ 
\frac{i \theta}{4 \pi \kappa \sin(\theta) }
\,.
%= \frac{i}{2 \pi \kappa}[ \sum_{j \in \mathbb{N}/2} \frac{e^{-i d_j 2 (\varphi - i \epsilon)}}{d_j} \chi_j(\theta) + \frac{\theta }{2 \sin(\theta)}]
\ee
The real part of this one-loop Feynman amplitude is given by a log, while the imaginary part satisfies the Euclidean version of the optical theorem; that is, explicitly for all $\theta,\vphi\in[0,2\pi]$:
%and gives overall for $\varphi \in [0, \pi]$:
\begin{equation}
\text{Re}\,   \Bell^{E}_\varphi(g) = 
\frac{-1}{4 \kappa \pi \sin\theta} \log\left|\frac{1 - \cos2 \varphi +\cos\theta - \cos(2 \varphi-\theta)}{1 - \cos2 \varphi +\cos\theta - \cos(2 \varphi+\theta)}\right|
\,,   
\qquad
   i\text{Im}\,\Bell^{E}_{\varphi}(p) +  \ell^{0, E}_\varphi(p) = \frac{\pi}{4  \kappa \sin\theta }
   \,.
\label{offshellSU(2)}
\end{equation}
Thus, with this corrected $\SU(2)$ Feynman propagator, the NCQFT verifies the Euclidean version of the Cutkosky cut rule  at one-loop. And, unlike the $\SO(3)$ case, there is no surprise when the mass angle  $2 \varphi$ goes above $\pi/4$.{We have thus cured the awkward feature at high mass of the $\SO(3)$ propagator, by avoiding the somewhat unphysical identification between momenta $\theta$ and $\pi-\theta$.}
This gives hope to get a truly unitary theory in  Lorentzian signature, at least at the one-loop level, by similarly correcting the Feynman propagator . We show in  the next section that it is indeed the case.

%%%%%%%%
\section{NCQFT Unitarity and Feynman Propagators: Lorentzian theory}
%%%%%%%%

In this part, we move to the Lorentzian quantum field theory, based on momenta living in the groups $\SO(2,1)$ or its double cover $\SU(1,1)\sim \SL(2,\R)$.  This NCQFT is meant to drive the propagation of quantum fields coupled to 3d quantum gravity in Lorentzian signature, described by the Lorentzian version of the Ponzano Regge state-sum \cite{Freidel:2000uq,Davids:2000kz}.
We define the Hadamard and Feynman propagators encoding the mass-shell conditions and off-shell correlations of the quantum field in non-commutative space-time. We then compute the one-loop integrals and show that the NCQFT based on $\SU(1,1)$ with a suitably 0-mode corrected Feynman propagator is indeed unitary (at the one-loop level).

%%%
\subsection{Lorentzian Feynman and Hadamard propagators}
%%%

In order to generalize our approach from the Euclidean case to the Lorentzian theory, we face four potential issues:
\begin{itemize}

\item properly deal with the non-compactness of the Lorentz groups, $\SO(2,1)$ and $\SU(1,1)$, with infinite-dimensional unitary representations;

\item carefully deal with the difference between time-like and space-like momenta, corresponding to elliptic and hyperboloic group elements having compact or non-compact conjugation orbits;

\item keep in mind the $\Z_{2}$ difference between $\SO(2,1)$ and $\SU(1,1)$, which ended up playing a crucial role in the Euclidean field theory;

\item understand the possible admissible corrections to the Feynman propagator, and use this ambiguity to identify the one(s) compatible with one-loop unitarity.

\end{itemize}

%We derive the expression of the propagators and then compute the associated one-loop corrections. In particular, we carefully distinguish two types of propagators, one taking value on elliptic element $g(\theta, \hat{u}), \; \hat{u}^{2} = 1$ and the other on hyperbolic one $g(t, \hat{a}), \; \hat{a}^{2} = -1$.

Indeed, the foremost difference between the Euclidean and Lorenztian cases, is that the Lorentz groupes $\SO(2,1)$ and $\SU(1,1)$ are non-compact compared to their Euclidean counterparts, $\SO(3)$ and $\SU(2)$. The direct consequence is that their unitary representations are automatically infinite-dimensional. This algebraic fact has a direct analytical consequence since, following the Plancherel formula, central functions and distributions are to be expanded on the characters of these unitary representations.

The unitary representations of $\SU(1,1)$ involved in the Plancherel decomposition are called the principal series \cite{Bargmann:1946me}. There are of two types (see e.g. \cite{Davids:2000kz,Kitaev:2017hnr} for more details):
\begin{itemize} 
\item The principal continuous series are parametrized by a real number $s>0$ and a parity $\varsigma=\pm$. Their Hilbert space is spanned by states $|(s,\varsigma), m\ra$ with $m\in\Z$ if $\varsigma=+$ or $m\in\Z+\f12$ if $\varsigma=-$. The eigenvalue of the Casimir $\vJ^{2}=J_{z}^{2}-K_{x}^{2}-K_{y}^{2}$  is $-(s^{2}+\f12)$ while the number $m$ diagonalizes $J_{z}$. The character of the representation does not depend on the parity $\varsigma$, its vanishes on elliptic group elements $g(\theta,\hu)$ whiile it oscillates on hyperbolic group elements:
\be
    \chi_s(\theta) = 0\,, \qquad \chi_s(t,\pm) = \frac{\cos\,st}{|\sinh t|} \,,
\ee
where one should remember that hyperbolic group elements are conjugate either to $(\cosh t+i\sinh t\ha\cdot\vec{\tau})$ or to $-(\cosh t+i\sinh t\ha\cdot\vec{\tau})$, without any possibility of switch sign by conjugation, even if a group element with boost parameter $t$ is obviously conjugated to with boost parameter $-t$ (but without changing the overall sign in front of the group element).

Representations of $\SO(2,1)$ are only the ones with positive parity, $\varsigma=+$.

\item The positive and negative discrete are parametrized by a half-integer $j$ and a sign $\eps=\pm$. Their Hilbert space is spanned by states $|(j,\eps), m\ra$ with $m\in j+1+\N$ if $\eps=+$ or $m\in-(j+1+\N)$ if $\eps=-$. The Casimir is given by $j(j+1)$. The character of the representation oscillates on elliptic group elements $g(\theta,\hu)$, it does not vanish on the hyperbolic group elements but admits an exponential tail:
 \be
   \chi_j^{\eps}(\theta) = -\eps \frac{e^{ i \,\eps d_j \theta}}{2i \sin \theta}
   \,, \qquad
   \chi_j^{\eps}(t,\pm) = (\pm)^{2j}\frac{e^{- d_j |t|}}{2|\sinh t|}
   \,.
 \ee
 Representations of $\SO(2,1)$ are only the ones with integer spin $j\in\N$, which satisfy $\chi_j^{\eps}(\theta) =\chi_j^{\eps}(\theta+\pi)$.

\end{itemize}
%
%And we define $\chi_j$ as the sum of $\chi_j^{+} $and $\chi_j^{-}$. On elliptic elements, $\chi_j(\theta) $ equals its Euclidean counterpart modulo a minus sign.
%
The Plancherel measures for $\SO(2,1)$ and $\SU(1,1)$  was found by Harish-Chandra \cite{Harish} and give the decomposition of the $\delta$-distribution in terms of the characters of the principal series of unitary representations.
Let us start with $\SU(1,1)$. The Plancherel formula is:
\be
\delta(g) = \sum_{j \in \f\N2} d_j \chi_j + \int \rd s \, \mu(s) \chi_s
\,,\quad\textrm{with}\,\,
\mu(s)
=
\frac{s}{2} \left[ \textrm{cotanh}\f{\pi s}2 +  \tanh\f{\pi s}2 \right]
%\frac{s}{2} \left[ \textrm{cotanh}(\pi s /2) +  \tanh(\pi s /2) \right]
\,,
\ee
where we write $\chi_{j}=\chi^{+}_j+\chi^{-}_{j}$. Notice the sign switch with respect to the Euclidean case:
\be
\chi_{j}(\theta)=\chi^{+}_j(\theta)+\chi^{-}_{j}(\theta)
=
-\f{\sin d_{j}\theta}{\sin\theta}
\,.
\ee

To impose the mass-shell condition, we need a distribution enforcing that the group element $g$ is elliptic and fixing its class angle to a given $\vphi\in[0,\pi]$. This is given by:
\be
\label{deltaphiL}
\delta_\varphi(g) =
%\frac{4 \tan \varphi}{\kappa^2}
\sum_{j \in \frac{\N}{2}} [\chi^{+}_j(\varphi) \chi^{-}_j(g) + \chi^{-}_j(\varphi) \chi^{+}_j(g)] + \int \mathrm{d} s f_s(\varphi) \chi_s(g)
\,.
\ee
The coefficients $f_{s}$ are determined by the condition that $\delta_\varphi(g)$ vanishes on hyperbolic group elements.
In fact, one can compute the sum over spins for a group element $g(t,+)$ as a geometric series:
\be
\sum_{j \in \frac{\N}{2}} [\chi^{+}_j(\varphi) \chi^{-}_j(t,+) + \chi^{-}_j(\varphi) \chi^{+}_j(t,+)] 
=
\f{-1}{2\sin\vphi|\sinh t|}\sum_{j\in\f\N2}e^{-d_{j}|t|}\sin d_{j}\vphi
%\nn\\
=
\f{-1}{4|\sinh t|}\f1{(\cosh t-\cos\vphi)}
\,.
\ee
Then the condition that $\delta_{\vphi}(t,+)=0$ imposes that the coefficients $f_{s}$ be the Fourier transform of this function in $t$:
\be
\int \mathrm{d} s f_s(\varphi) \cos st
=
\f{1}{4(\cosh t-\cos\vphi)}\,,
\ee
which, by a straightforward residue computation,\footnotemark{}
\footnotetext{To apply residue formula, we chose a rectangle un the upper plane of length $ L$ and height $2 \pi$, along vertical segements we can easily show that the integral of $\phi(z) = \frac{e^{i z s}}{ 4(\cosh z - \cos \varphi)}$ vanish. It remains the two horizontal segements which give a non zero contribution to the function:
\begin{equation}
    f_s(\varphi) = \frac{1}{(1 - e^{- 2 \pi \varphi)}} \oint_{\text{rect}} \varphi(z) dz 
\end{equation}.
In the rectangle there are two poles, $i \varphi$ and $i (2 \pi - \varphi)$, using the residue theorem:
\begin{equation}
    f_s(\varphi) = \frac{2 \pi i}{1 - e^{-2 \pi s}}[\lim_{z\to i \varphi } (z - i \varphi) \phi(z) + \lim_{z\to i( 2 \pi - \varphi)} (z - i (2 \pi -\varphi)) \phi(z)]
\end{equation}
Finally thanks to L'Hôpital's rule one gets \ref{coeffDiracfs}
}
finally yields for arbitrary $\vphi\in[0,2\pi]$:
\be
\label{coeffDiracfs}
f_s(\varphi)
=
\frac{1}{2 \sin{\varphi}} \frac{\sinh(\pi-\varphi) s}{\sinh\pi s}
\,.
\ee
%
%\footnotetext{
%{\bf TYPE FOOTNOTE}
%}
%
Then evaluating the same expression on elliptic group elements $g(\theta,\hu)$ gives:
\be
\delta_\varphi(g) = \frac{1}{4 \sin\varphi\, \sin\theta} \Big{[}2 \pi \delta(\theta - \varphi) - 1\Big{]}
\,.
\ee

We can adapt all these formulas to the $\SO(2,1)$ case. One should remember that we need to identify $g$ and $-g$, or equivalently identify the angles $\theta$ and $\theta+\pi$. Then the Plancherel formula becomes
\be
\tdelta(g) = \sum_{j \in \N} d_j \chi_j + \int \rd s \, \tmu(s) \chi_s
\,,\quad\textrm{with}\,\,
\tmu(s) =\frac{s}{2}\, \tanh(\pi s /2)
\,.
\ee
Then fixing the mass-shell  $\SO(2,1)$ amounts to fixing the angle $\theta$ to either  $\vphi$ or $\vphi+\pi$,
%the $\SO(2,1)$ Hadamard propagator, fixing the mass-shell to $\theta=\vphi$ $\theta=\vphi+\pi$, is given by:
\begin{equation}
\tdelta_\varphi(g)
%\delta_\varphi^{\SO(2,1)}(g)
=
%\frac{\cos \varphi}{2 }
\f12
\Big{[}\delta_\varphi(g) + \delta_{\varphi + \pi}(g)\Big{]}
=
%\frac{4 \sin \varphi}{\kappa^2}
\sum_{j \in  \mathbb{N}} [\chi^{+}_j(\varphi) \chi^{-}_j(g) + \chi^{-}_j(\varphi) \chi^{+}_j(g)] + \int \mathrm{d} s F_s(\varphi) \chi_s(g) 
\,,
\end{equation}
where the coefficients $F_{s}$ are derived from the $f_{s}$'s:
%by the condition that $\tdelta_\varphi(g)$ vanishes on hyperbolic group elements\footnotemark:
\be
\label{coeffDiracFs}
F_s(\varphi)
=
\f12\Big{[}f_s(\varphi)+f_s(\varphi+\pi)\Big{]}
=
\frac{1}{4 \sin\vphi} \frac{\cosh s(\f\pi2 - \varphi)}{\cosh \f\pi2s }
\,.
\ee
%
%\footnotetext{
%Assuming $F_s$ is even in $s$, it must satisfy:
%\begin{equation*}
%\frac{1}{2|\sinh t|} \int \rd s  e^{ist}F_s(\varphi)
%=
%- \sum_{j \in  \mathbb{N}} [\chi^{+}_j(\varphi) \chi^{-}_j(g) + \chi^{-}_j(\varphi) \chi^{+}_j(g)] 
%=
%\frac{-1}{2 \sin\varphi |\sinh t|}\sum_{j \in \mathbb{N}} e^{-i d_j|t|} \sin\varphi
%\,.
%\end{equation*}
%We can first evaluate the sum:
%%
%$\frac{1}{\sin\varphi} \sum_{j \in \N} e^{-i d_j|t|}(\sin d_j \varphi )= \frac{- \cosh(t)}{2(\sinh^{2}t + \sin^{2}\varphi)}$
%%
%and then compute in the complex plane:
%\begin{equation*}
%\int_{\R} \rd t\,e^{ist} \frac{ \cosh t}{\sinh^{2}t + \sin^{2}\varphi}
%=
%\frac{\pi}{\sin\varphi} \frac{\cosh s(\f\pi2 - \varphi)}{\cosh \f\pi2 s }
%\,, \qquad
%\frac{ \cosh t}{\sinh^{2}t + \sin^{2}\varphi} = \frac{1}{\sin \varphi}
%\int_0^{\infty} \rd s \, \frac{\cosh s(\f\pi 2 - \varphi)}{\cosh\f\pi2 s}
%    \,,
%\end{equation*}
%recovering the wanted expression.
%}
%
This distribution vanishes on hyperbolic group elements by construction and yields the expected $\delta$-distribution on $\SO(2,1)$:
%One can explicitly sum this expression to check that one indeed gets:
\be
\tdelta(g)
=
\frac{\pi}{ 4 \sin\theta\, \sin\varphi} \Big{[}\delta(\theta - \varphi) - \delta(\theta - \varphi- \pi)\Big{]}
\,.
\ee
The interested reader will find more details on distributions on $\SO(2,1)$ and $\SU(1,1)$ in appendix \ref{app:distriL}.

\medskip

Let us now turn to the Feynman propagator. We read it off directly from our field theory action \eqref{actionG} \cite{Freidel:2005bb},
%We now turn to the Feynman propagator, which is defined the same way as the Euclidean theory \cite{Freidel:2002xb}:
\begin{equation}
K^{\rL}_\varphi(g)
=
\frac{1}{P^{2}(g) - \kappa^{2} \sin^{2}(\varphi) + i \epsilon}
=
\left\{
\begin{array}{l}
\displaystyle\frac{1}{\kappa^{2} (\sin^{2}\theta - \sin^{2}\varphi + i \epsilon)}\,,
\vspace*{2mm}\\
\displaystyle\frac{-1}{\kappa^{2}(\sinh^{2}t + \sin^{2}\varphi)}\,.
\end{array}
\right.
%\frac{i}{\kappa^{2} (\sin^{2}(\theta) - \sin^{2}(\varphi) + i \epsilon)} \quad \text{or} \quad = \frac{-1}{\kappa^{2}(\sinh^{2}(t) + \sin^{2}(\varphi))}
\end{equation}
This propagator is the same as in \cite{Imai:2000kq,Sasai:2009jm} but with a different sign convention due to a difference of signature $(-,+,+)$.
A straightforward calculation gives the character decomposition of $K_\varphi^{\rL}$: 
%\footnote{We first check the formula on elliptic elements: $
%    - \sum_{j \in \mathbb{N} } \chi_j(\theta) \frac{2 e^{- i d_j (\varphi - i \epsilon)}}{\cos(\varphi)} = \frac{1}{\sin^{2}(\theta) - \sin^{2}(\varphi) + i \epsilon} $ . Then, in the case of a hyperbolic group element, se get for $t > 0$: \begin{equation*}
%        \frac{-1}{\sinh^{2}(t) + \sin^{2}(\varphi)} + \sum_{j \in \mathbb{j}} \chi_j(t)  \frac{2 e^{- i d_j (\varphi - i \epsilon)}}{\cos(\varphi)} = \frac{-i \sin(\varphi)}{\cos(\varphi)} \frac{\cosh(t) }{\sinh(t) (\sinh^{2}(t) + \sin^{2}(\varphi))} 
%    \end{equation*} And one recognize the explicit expression of $\int ds \; F_s(\varphi) \chi_s(t)$ derived previously}
%
\be
K^{\rL}_{\varphi}(g)
=
\frac{-4i \tan\varphi}{\kappa^{2}}
\left[
\sum_{j \in \mathbb{N}} \chi_j^{-}(\varphi - i \epsilon) \chi_j(g) + \int \rd s \, F_s(\varphi) \chi_s(g)
\right]
\,.
\ee
We recognize in the first term the  the character expansion of the Euclidean Feynman propagator. In contrast, the second term gives the exponentially decreasing tail of the Feynman propagator for space-like intervals. This second term was missed in \cite{Freidel:2005bb}.

We deduce the  Hadamard propagator from the imaginary part of this Feynman propagator:
%We recover the $\SO(2,1)$  Hadamard propagator from this Feynman propagator:
%Finally, we compute the Hadamard propagator, built with our previous Dirac distribution, considering the two sheets of hyperboloid:
\begin{equation}
\cH^{L}_{\vphi}(g)\equiv
i K_\varphi^{\rL}(g) - i K_{-\varphi}^{\rL}(g) 
=
\frac{4 \tan\varphi}{\kappa^{2}}
\tdelta_{\vphi}(g)\,,
%\delta_\varphi^{H}(g)
%\quad \text{with} \quad \delta_\varphi^{H}(g) = \delta_\varphi^{W+}(g) + \delta_\varphi^{W-}(g)
\end{equation}
which gives the $\SO(2,1)$ mass-shell condition. We thus refer to $\cH^{L}_{\vphi}$ as the $\SO(2,1)$ Hadamard propagator.
The reason why this leads to $\SO(2,1)$ and not to the $\SU(1,1)$ mass-shell is the same as the Euclidean case. Indeed, the Feynman propagator $K_\varphi^{\rL}(g)$ does not distinguish the four group elements $\pm g, \, \pm g^{-1}$. The symmetry of the propagators under the exchange $g \leftrightarrow g^{-1}$ is natural and corresponds to the standard $\vec{p} \leftrightarrow - \vec{p}$ symmetry. What's more subtle is the confusion $g \leftrightarrow -g$, which does not have any counterpart in the standard commutative field theory.
Similarly to what we achieved in Euclidean signature, we can lift this confusion by introducing an appropriate $\SU(1,1)$ propagator.
%
%This Hadamard propagator does not distinguish the four group elements $\pm g, \; \pm g^{-1}$. The sign "confusion" might be avoided by using the full structure of SU(1,1). At the same time, it is natural and expected that the Hadamard propagator is symmetric under the exchange $g \longleftrightarrow g^{-1}$ (corresponding to the usual $\vec{p} \longleftrightarrow - \vec{p}$ symmetry). The character expansion of the various propagators is close to the one guessed in \cite{Freidel:2005bb} but for the tail on continuous representations and the distinction between summing on $j \in \mathbb{N}$ or $j \in \mathbb{N}/2$. \\

Indeed, let us now define the $\SU(1,1)$ Feynman propagator, by including odd parity representations (i.e. from the continuous series with $\varsigma=-$ and from the discrete series with half-integer spins), as well as including an extra 0-mode:
\be
\mathbf{k}_\varphi^{\rL}(g)
=
k_\varphi^{\rL}(g)-\f1{\ka^{2}}\chi_{-\f12}(g)
\,,
\qquad\textrm{with}\quad
k_\varphi^{\rL}(g)
=
\frac{-2 }{\kappa^{2}} \sum_{j \in \f\N2 } e^{-i d_j (\varphi - i \epsilon)} \chi_j(g)
- \frac{4i \sin\varphi}{\kappa^{2}} \int_{0}^{+\infty} \rd s \, f_s(\varphi) \chi_s(g)
\,,
\ee
where we included an extra-mode,  similarly to what we did in the Euclidean case:
\be
\chi_{-\f12}(\theta) = \frac{-1}{ \sin\theta}
\,,\qquad
\chi_{-\f12}(t) = \frac{1}{ \sinh t}
\,.
\ee
This is an eigenvector of the $\su(1,1)$ Casimir with eigenvalue $-\f14$, thereby formally corresponding to a spin $j=-\f12$, thus the notation. It should correspond to  the character of the exceptional (or mock) discrete representation, which is the limit of the principal discrete series of unitary representations of $\SU(1,1)$ and has vanishing Plancherel measure.

The integral over the continuous series labeled by $s$ was already computed above, while the series in the spin $j$ can be resummed exactly as a geometric series for $\eps>0$. The convergence is numerically very fast, except close to $t\rightarrow 0^{+}$, and gives:
\begin{equation}
\bk_\varphi^{\rL}(\theta)
= 
\frac{1}{\kappa^{2}\left(\cos\vphi-\cos\theta + i \epsilon\sin\vphi \right)}
+
\frac{1}{\kappa^{2} \sin\theta}
\,,\qquad
\bk_\varphi(t)
= 
\frac{-1}{\kappa^{2} (\cosh t - \cos\varphi)}
\,.
\end{equation}
This Feynman propagator leads a $\SU(1,1)$ Hadamard propagator which properly reflects the mass-shell condition $\theta=\pm\vphi$, where the sum over both signs is due future and past orientation for the time-like momentum: 
\begin{equation}
\sh^{L}_{\vphi}(g)
\equiv
i \bk_\varphi^{\rL}(g) - i\bk_{-\varphi}^{\rL}(g)
=
\frac{4 \sin\varphi}{\kappa^{2}}
 \deltaH_\varphi(g)
\,, \qquad \text{with} \quad
\deltaH_\varphi(g)= \delta_\varphi(g) + \delta_{-\varphi}(g)
\,.
\end{equation}
We show below that this $\SU(1,1)$ propagator defines a unitary quantum theory at the one-loop level.

%%%
\subsection{Lorentzian one-loops integrals and unitarity}
%%%

Now that the propagators of the Lorentzian field theory have well-defined as distributions on the Lie groups $\SO(2,1)$ and $\SU(1,1)$, we can analyze the one-loop diagrams of the quantum field theory and check whether they satisfy the optical theorem or violates unitarity.
We consider the one-loop diagram with incoming group element $g(\theta, \hat{u})\in\SU(1,1)$. To distinguish past and future orientations, one must restrict the angle $\theta$ to lay in the interval $[0,\pi]$.

First, let us analyze the $\SO(2,1)$ theory with its Feynman propagator $K_\varphi^{\rL}(g)$.  Due to the identification between the group elements $g$ and $-g$ in the case of $\SO(2,1)$, we further restrict  the range of the angle  to $\theta\in[0, \pi/2]$.
The off-shell and on-shell one-loop integrals read:
\begin{equation}
\cL_{\varphi}^{\rL}(g ) 
= 
 \frac{\kappa^{2}}{8 \pi} \int dh K_\varphi^{\rL}(h) K_\varphi^{\rL}(hg)
\,, \qquad
\cL_{\varphi}^{0,\rL}(g )
=
i \frac{\kappa^{2}}{16 \pi}  \int \rd h\, \tdelta_\varphi(h)  \tdelta_\varphi(hg)
\,.
\end{equation}

To compute the off-shell one-loop in the  non-commutative Lorentzian theory, we look at the non-deformed field theory and notice that the Euclidean and Lorentzian off-shell one-loop amplitudes are similar and involve the same integrals.
This feature is carried to the deformed theory, and we actually show  that Lorentzian off-shell computation is equal to Euclidean off-shell one-loop.

Indeed, in the Euclidean case, we consider $\SO(3)$ group elements, the external momentum  $g(\theta, \hat{u})$ and the momentum along the loop $h(\alpha, \hat{v})$, and call $\lambda$ the class angle of $hg \in \SO(3)$. The $\SO(3)$ off-shell one loop  then explicitly reads (see Appendix \ref{appendixA} for details): 
\begin{equation}
\mathcal{L}^{\rE}_\varphi(g)= \frac{i \kappa^{3}}{8 \pi^{2}} \int^{ \pi}_0 \sin{\alpha}^{2} d \alpha \int_0^{\pi} \sin{\beta} \; d \beta K_\varphi(h) K_\varphi(hg)= \frac{i \kappa^{3}}{8 \pi^{2}} \int_0^{ \pi} \sin{\alpha} d \alpha \int_{ \alpha + \theta}^{ \alpha - \theta} \frac{\sin{\lambda}}{\sin{\theta}} d \lambda K_\varphi(h) K_\varphi(hg) 
\,,
\end{equation}
where $\beta$ is the angle between the rotation axis of $g$ and $h$.
%where we have used relation \ref{EuclidAngleRelation} between the angle $\theta, \alpha, \beta$ and $\lambda$ to performed variable change.  
%
In the Lorentzian case, we consider $\SO(2,1)$ group elements, the past-oriented external momentum $g(\theta, \hat{u}) \in \cH^{-} $ and the loop momentum $h(\alpha, \hat{v}) \in \SO(2,1)$. We call $\lambda $ the angle of the composed momentum $hg \in \SO(2,1)$. After making the suitable change of variables from the boost parameter between $g$ and $h$ to the rapidity of the composed momentum, the off-shell one loop reads (see appendix \ref{appendixA} for details):
%\begin{equation}
%\begin{split}
%    \mathcal{L}^{\rL}_\varphi(g) & = \frac{i \kappa^{3}}{8 \pi}  \int_0^{\pi} \sin{\alpha}^{2} d \alpha [\int_{\cH^{+}} \frac{d \hat{u}}{4 \pi} + \int_{\cH^{-}} \frac{d \hat{u}}{4 \pi} ] K_\varphi(h) K_\varphi(hg) \\
%    & = \frac{i \kappa^{3}}{8 \pi^{2}}  \int_0^{\pi} \sin{\alpha}^{2} d \alpha [\int_{0, \cH^{+}}^\infty \sinh{s} ds +  \int_{0, \cH^{-}}^\infty \sinh{r} dr ] K_\varphi(h) K_\varphi(hg) 
%\end{split}
%\end{equation}
%As before, we need to make a variable change using relations between $s,r$ and angles:
%\begin{equation}
%    \begin{split}
%   & \cos(\lambda_\pm) = \cos(\theta) \cos(\alpha) - \sin(\theta) \sin(\alpha) ( \cosh(s))\\
%    &\cos(\lambda_\pm) = \cos(\theta) \cos(\alpha) - \sin(\theta) \sin(\alpha) (- \cosh(r))
%    \end{split}
%\end{equation}
%After variable change we get:
%\begin{equation}
%\begin{split}
%     \mathcal{L}^{\rL}_\varphi(g) &  = \frac{i \kappa^{3}}{8 \pi^{2}} \int_0^{ \pi} \sin{\alpha} d \alpha [\int^0_{ \alpha - \theta} \frac{- \sin{\lambda}}{\sin{\theta}} d \lambda + \int^\pi_{ \alpha + \theta} \frac{\sin{\lambda}}{\sin{\theta}} d \lambda ] K_\varphi(h) K_\varphi(hg)  =  \mathcal{L}^{\rE}_\varphi(g)
%\end{split}
%\end{equation}
%
\beq
\mathcal{L}^{\rL}_\varphi(g)
& = &
\frac{i \kappa^{3}}{8 \pi}  \int_0^{\pi} \sin{\alpha}^{2} \rd \alpha
\int_{\cH^{+}\cup\cH^{-}} \frac{\rd \hat{u}}{4 \pi}  K_\varphi(h) K_\varphi(hg)
\nn\\
&=&\frac{i \kappa^{3}}{8 \pi^{2}} \int_0^{ \pi} \sin{\alpha} \rd \alpha
\bigg{[}\int^0_{ \alpha - \theta} \frac{- \sin{\lambda}}{\sin{\theta}} d \lambda + \int^\pi_{ \alpha + \theta} \frac{\sin{\lambda}}{\sin{\theta}} \rd \lambda
\bigg{]}
 K_\varphi(h) K_\varphi(hg)
\nn\\
&=&  \mathcal{L}^{\rE}_\varphi(g)
\,.
\eeq
This shows that the Lorentzian expression is equal to its Euclidean counterpart, which has already been computed before.
This is consistent with the standard undeformed QFT where Euclidean and Lorentzian off-shell one-loops were also equal. 
This yields:
\begin{equation}
\cL^{\rL}_\varphi(g) = \left\{
\begin{array}{ll}
\displaystyle\frac{- i}{ 8 \pi \kappa \sin{\theta} \cos^{2}{\varphi}} \log{\left(\frac{\sin{2 \varphi} - \sin{\theta}}{\sin{2 \varphi} + \sin{\theta}}\right)}\,,
& \qquad \text{if}\quad\theta \in [0, 2 \varphi]
\,,
\vspace*{1mm}\\
\displaystyle\frac{- i}{ 8 \pi \kappa \sin{\theta} \cos^{2}{\varphi}}\log{\left(\frac{- \sin{2 \varphi} + \sin{\theta}}{\sin{2 \varphi} + \sin{\theta}}\right)} + \frac{i}{8 \kappa \sin{\theta} \cos{\varphi}^{2}}
\,,
&\qquad \text{if}\quad \theta \in [2 \varphi, \pi/2] \,.
\end{array}
\right.
\end{equation}

As for the on-shell amplitude, one can  integrate exactly the $\delta$-distribution entering the on-shell loop diagram  (see appendix \ref{appendixC} for details).
For $\theta \in [0, \pi/2] $ and $\varphi \in [0, \pi/4]$, we get:
\begin{equation}
\textrm{for }\varphi \in [0, \pi/4]\,,\qquad
    \cL_{\varphi}^{0, \rL}(g) = \left\{
    \begin{array}{ll}
         0 &  \text{if } \theta \in [0, 2 \varphi] \\
        \displaystyle\frac{i}{8  \kappa \cos{\varphi}^{2} \sin{\theta}} &  \text{if } \theta \in [2 \varphi, \pi / 2]
    \end{array}
\right.
\end{equation}
%In the limit $\kappa \to \infty$, with incoming momentum $|\vec{P}| = \kappa \sin(\theta)$ and mass $M = \kappa \sin(\varphi)$, we recover the standard non-deformed  one-loop integral.
Thus for a mass angle $\varphi$ lying in the interval $[0,\pi/4]$, the cutting rule is satisfied.
However, when the mass $\varphi$ increases and lays in range $[ \pi/4, \pi/2 ]$, the on-shell one-loop integral always vanishes. 
\begin{eqnarray}
\textrm{for }\varphi \in [ \pi/4, \pi/2 ]\,,\qquad
    \cL_\varphi^{0, \rL}(g) = 0\,.
\end{eqnarray}
The cutting rule is no longer satisfied. This is the same issue that we encountered in the Euclidean theory at large mass.
Thus, we can conclude that the optical theorem only applies to angles $\varphi \in [0,\pi/4]$, corresponding to small masses $M\le\ka/\sqrt{2}$ (or bare mass $m\le \pi\ka/4$): 
increasing the mass fully puts in light the un-classical behavior of the on-shell one-loop amplitude. We recover exactly the result previously found \cite{Sasai:2009jm}, which showed the non-unitarity of the  3d Lorentzian  noncommutative scalar field theory based on the $\SO(2,1)$ momentum space.
Here, we go further that this previous work and show how to fix the unitarity of the NCQFT by using a Feynman propagator fully reflecting the $\SU(1,1)$ structure of the momentum space.

\bigskip

Let us then turn to the $\SU(1,1)$-based theory and its Feynman propagator $\bk_\varphi^{\rL}(g)$, involving all even and odd representations of $\SU(1,1)$ as well as the extra 0-mode coming from the exceptional discrete representation of spin $j=-\f12$. Working with the full $\SU(1,1)$ group, the angles can be restricted to the range $[0, \pi]$ in order to avoid redundancies. However, we drop the identification between $g $ and $-g$ defining the $\SO(2,1)$ subgroup, so the interval can not be reduced to $[0, \pi/2]$, we need to take $\theta, \varphi \in [0, \pi]$.
Following what we obtained for the Eculidean theory, we expect to find a cutting rule verified for all masses $\varphi$.

Off-shell and on-shell loop expressions are given by:
\begin{equation}
     \ell_\varphi^{\rL}(g) = \frac{i \kappa^{3}}{8 \pi} \int dh \bk^{\rL}_\varphi(h) \bk^{\rL}_\varphi(hg) + \frac{i \theta}{4 \pi \kappa \sin{\theta}}
     \,,\qquad
     \ell_\varphi^{0,\rL}(g) = \frac{i \kappa^{3}}{16 \pi} \int dh {\deltaH_\varphi}(h) {\deltaH_\varphi}(hg)
     \,.
\end{equation}
The $\SU(1,1)$ off-shell amplitude has a similar expression to its Euclidean counterpart, the $\SU(2)$ off-shell one loop. Details of the computation of both on-shell and off-shell integrals can be found in appendix \ref{appendixC}. The final result, for $\theta \in [0, \pi]$ and $\varphi \in [0, \pi]$, reads: 
\begin{equation}
    \ell_\varphi^{\rL}(g) = \frac{-1}{4 \pi \kappa \sin{\theta}} \log \Big( | \frac{1 - \cos{2 \varphi} + \cos{\theta} - \cos{2 \varphi - \theta}}{1 - \cos{2 \varphi} + \cos{\theta} - \cos{2 \varphi + \theta}} | \Big) + \left\{
    \begin{array}{ll}
         0 & \text{ if } \theta \in [0, 2 \varphi]\vspace*{2mm}\\
\displaystyle        \frac{i}{4 \kappa \sin{\theta}} & \text{ if }  \theta \in [2\varphi, \pi]
    \end{array}
\right.
\label{equ:SU(1,1)offshell}
\end{equation}
\begin{equation}
    \ell_{\varphi}^{0, \rL}(g) = \left\{
    \begin{array}{ll}
         0 & \text{si } \theta \in [0, 2 \varphi]\vspace*{2mm} \\
        \displaystyle\frac{i}{4  \kappa \sin{\theta}} & \text{sinon } \theta \in [2 \varphi, \pi ]
    \end{array}
\right.
\label{equ:SU(1,1)onshell}
\end{equation}
Thus, using the modified $\SU(1,1)$ Feynman propagator, we recover the Cutkosky rule for an arbitrary mass $ \varphi \in [0,\pi]$. The $\SU(1,1)$ noncommutative field theory is therefore unitary at the one-loop level, and can be considered as a legitimate physical field theory.
%for when the deformed momentum space is associated to SU(1,1) for elliptic element.

%%%%%%%%
\section*{Conclusion and Outlook}
\label{sec:dicussion}
%%%%%%%%

In this paper, we have looked into the unitarity of non-commutative quantum field theory (NCQFT) based on the $\R^3_{\ka}$ space-time in both Euclidean and Lorentzian signature. The non-commutativity of the space-time geometry comes from using a curved momentum space, the Lie groups $\SO(3)$ or its double-cover $\SU(2)$ in Euclidean signature, and $\SO(2,1)$ or its double-cover $\SU(1,1)\sim\SL(2,\R)$ in Lorentzian signature. These NCQFTs are understood to describe the effective dynamics of a scalar matter field coupled to 3d quantum gravity after integration over the geometrical degrees of freedom \cite{Freidel:2005me}. The unitarity of those effective field theories is therefore crucial, for two main points. First, as an effective theory, we would like to ensure that we have properly integrated over the gravitational sector. Second, as non-commutative field theory, we would like to make sure that these non-commutative models, which can legitimately stand on their own without mentioning quantum gravity, are well-defined mathematically and physically as quantum field theories.

Here, we analyzed the unitarity of these NCQFTs at the one-loop level, by checking the validity of the Cutkosky cut rules relating the off-shell one-loop amplitudes defined using the Feynman propagator to the on-shell one-loop integrals defined using the Hadamard propagator. Unitarity is a well-defined concept only in Lorentzian signature, but one can write the equivalent cut rule identity that mimics its standard Lorentzian counterpart. Previous work focussed on the $\SO(2,1)$-based theory \cite{Imai:2000kq,Sasai:2009jm} and found that the theory failed to be unitary when the scalar field mass was too high, the threshold being $\sin(\pi/4)$ times the Planck mass . Here we extended this previous analysis and found that, indeed, the $\SO(3)$ Euclidean NCQFT and the $\SO(2,1)$ Lorentzian NCQFT break the cut rule. On the other hand, we show that the NCQFTs based on the $\SU(2)$ and  $\SU(1,1)$ momentum spaces satisfy the cut rule and are therefore unitary at the one-loop level. This remedies the issues raised in \cite{Imai:2000kq,Sasai:2009jm} and sets the $\SU(1,1)$ NCQFT as a legitimate well-defined physical model.

It is important to stress the difference between a $\SU(2)$ momentum space and a $\SO(3)$ momentum space in Euclidean signature, and between a $\SU(1,1)$ momentum space and a $\SO(2,1)$ momentum space in Lorentzian signature. Focussing on the Lorentzian signature, since the signature does not affect the discussion, the mass-shell condition does not change and, thus, the Hadamard propagator are exactly the same whether we use $\SU(1,1)$  or $\SO(2,1)$. On the other hand, the Feynman propagator does change. And we identify the correct Feynman propagator for $\SU(1,1)$  to ensure one-loop unitarity. As one remembers that the Feynman propagator is directly the inverse of the kinematical differential operator in the quadratic term of the Lagrangian, this means that we are in fact modifying the field theory action. More precisely, instead of using a spin-1
character to define the mass-shell condition, i.e. the equivalent of $\vec{p}^{2}$, we use a spin-$\f12$ character, which would be the equivalent of an awkward $\sqrt{1-\vec{p}^{2}/\ka^{2}}$ with the Planck mass $\ka$. While this might look artificial when formulated in terms of classical momentum $\vec{p}$, it is actually very natural from the Lie group perspective. Such a kinematical operator was actually anticipated from the point of view of the group field theory approach to quantum gravity \cite{Fairbairn_2007}.

On top of this natural modification, which boils down to using half-integers spins instead of merely the integer spin representations, we further have to add an extra 0-mode to the Feynman propagator. This is a surprising feature. A similar term was discussed in early work on effective matter theories arising from the Ponzano-Regge model for 3d quantum gravity \cite{Oriti:2006wq}. In the purpose of imposing causality by enforcing a strict light-cone despite the non-commutativity of space-time coordinates, that work found some extra higher power resonance for vanishing momentum. Our extra term is also a massless resonance, thus a 0-mode, but is mathematically different from that previous proposal. It amounts to adding a spin $-\f12$ term to the Feynman propagator, corresponding to ``null'' representation of $\SU(1,1)$, i.e. the limit representation from the discrete series of representation, with vanishing Plancherel measure. 

Not only this new feature requires to revisit the non-commutative field equation and understand the physics behind it, and how it is reflected in the quantum field correlations, but it also underlines the potential need to revisit the role of vanishing Plancherel measure representation in spinfoam path integrals. Indeed such representations were naturally cast aside, since there were not involved in the Fourier modes over the gauge groups in the standard construction of spinfoams (see e.g. \cite{livine2024spinfoam} for a recent overview). Moreover, since the necessary representation represents the quantum equivalent of null vectors, this hints towards the potentially important role that null structures could play in spinfoam models (see e.g. \cite{Speziale:2013ifa}), echoing the recent works on null and Carrollian structures in general relavity e.g. \cite{Ciambelli:2023mir,Odak:2023pga}.

Finally, we point out that  our study of unitarity was performed only at the one-loop level. To our knowledge,  no study has investigated, up to now,  unitarity at higher order in noncommutative geometry.
Renormalizability was more studied, with great results on the renormalization at all orders of the  $\phi^{4}$ scalar field with Moyal non-commutativity in two dimensions \cite{Grosse_2003} and four dimensions \cite{Grosse_2005}.
We hope that, similarly, our one-loop analysis will extend to all orders and that the $\SU(1,1)$ NCQFT on $\R^{3}_{\ka}$ will indeed be fully unitarity.
%One could expect to find an extension of unitarity at all orders; it should be examined in the future. However, the result of this paper is a promising first step in studying unitarity in the context of spinfoam.
%
This would confirm that non-commutative field theories, based on curved momentum spaces, as good viable candidates for effectively describing  matter fields coupled  to quantum gravity.

%%%%%%%%%%%%%%%%%%%%%%%%%%%%%

\section*{Acknowledgements}
EL is grateful to Laurent Freidel for suggesting to study the unitarity of non-commutative field theories back in 2005.

%%%%%%%%%%%%%%%%%%%%%%%%%%%%%
\appendix
%\addtocontents{toc}{\protect\setcounter{tocdepth}{-1}}

%%%%%%%%
\section{Measures on Lie groups}
%%%%%%%%
\label{appendixA}
We parameterize SU(2) using its spinorial representation. A group element $g$ is defined by its angle $\theta \in  [0,2\pi]$ and a unit vector $\hat{u} \in \cS^{2}$. The normalized Haar measure is:
\begin{equation}
    \int_{SU(2)} \rd g = \int_0^{2\pi} \;\sin^{2}\theta  \rd\theta \int_{\cS^{2}} \frac{\rd^{2} \hat{u}}{(2 \pi)^{2}} = \frac{1}{4 \pi^{2}}\int_0^{2\pi} {\sin^{2}\theta} \rd\theta \int_0^{\pi} \sin\beta \; \rd \beta  \int_0^{2 \pi} \rd \gamma 
    \label{measureSU(2)}
\end{equation}
where $\beta$ and $\gamma$ are the Euler angle for $\hat{u} = (\sin\beta \cos\gamma, \sin\beta \sin\gamma, \cos\beta)$. Note that the group element $(\theta, \hat{u})$ and $(\theta+ \pi, -\hat{u})$ are identified. Therefore, we can restrict the range of $\theta$-integration to $[0,\pi] $ and add a factor $2$ to keep the measure normalized. \\
To link with the deformed measure on momentum, we start with the momentum $\vec{P}(g) = \kappa \sin(\theta) \hat{u}$. Computing the Jacobian du to variables change $(P_1,P_2,P_3) \to \kappa( \sin\theta \sin\beta \cos\gamma, \sin\beta \sin\gamma, \cos\beta)$, we can show that:
\begin{equation}
    \frac{1}{\pi^{2} \kappa^{3}} \int_{|P|< \kappa} \frac{d^{3} \vec{P}}{\sqrt{1 - \frac{\vec{P}^{2}}{\kappa^{2}}}} = \frac{1}{4 \pi^{2}}\int_0^{2\pi} \sin^{2}\theta \; \rd\theta \int_0^{\pi} \sin\beta \; \rd \beta  \int_0^{2 \pi} d \gamma = \int_{SU(2)} \rd g
\end{equation}
In Lorentzian space, the situation is slightly more complicated due to the noncompactness of the Lie group SU(1,1) (or SO(2,1)). A suitable measure has been derived in \cite{Freidel:2002xb}; the idea is to separate integration on elliptic and hyperbolic elements. A function $f$ can be described by its action on both Cartan subgroups (the subgroup of rotations and the subgroup of boosts) of SU(1,1), $f = f_0 + f_1$, where $f_0$ (resp. $f_1$) as support on elliptic (resp. hyperbolic) elements and which can be conjugated to rotation (resp. boosts). Formulas for the measure read:
\begin{equation}
    \rd \mu_{SU(1,1)}(f_0) = \frac{1}{\pi} \int_0^{ 2 \pi} \sin^{2}{\theta} \rd \theta
\int_{\cH^{+}\cup\cH^{-}} \frac{\rd^{2} \hat{u}}{4 \pi} \,
%(\int_{\cH^{+}} \frac{\rd^{2} \hat{u}}{4 \pi} + \int_{\cH^{-}} \frac{\rd^{2} \hat{u}}{4 \pi} )
 f_0(\theta, \hat{u})
 \,,
 \quad \quad  \rd \mu_{SU(1,1)}(f_1) = \frac{1}{\pi} \int_0^{\infty} \sinh^{2}t \;
 \rd t \int_{\cH} \frac{\rd^{2} \hat{a}}{4 \pi} f_1(t, \hat{a}) 
 \,,
\end{equation}
where $\cH^\pm$ are the two sheets hyperbola, $\cH$ the one sheet hyperboloid, $\hat{u}, \hat{a}$ units vectors of elliptic and hyperbolic elements, $\hat{u} \in \cH^{\pm}, \hat{a} \in \cH$. We can find an explicit parametrization of those elements and integration over the two and one-sheet hyperbola, $\hat{u} = (\sinh(s) \cos(\gamma), \sinh(s) \sin(\gamma), \pm \cosh(s))$, with $\pm$ depending of $\hat{u} $ in lower or upper hyperboloid and $\hat{a} = (\cosh(s) \cos(\gamma), \cosh(s) \sin(\gamma), \pm \sinh(s)) $. With those expressions, measures on hyperbola are given by: 
\begin{equation}
        \int_{\cH^{\pm}} \frac{\rd^{2} \hat{u}}{4\pi^{2}}
        = \frac{1}{4 \pi^{2}}\int_{0}^{\infty} \sinh s \; \rd s \int_0^{2\pi} \rd \gamma
        \,,\qquad
        \int_{\cH} \frac{\rd^{2} \hat{a}}{4\pi^{2}} = \frac{1}{4 \pi^{2}}\int_{-\infty}^{\infty} \cosh s \; \rd s \int_0^{2\pi} \rd \gamma \label{measureLorentz}
\end{equation}

Finally, as in the Euclidean case, we can make a variable change, with $\vec{P}(g) = \kappa \sin\theta \,\hat{u}$ or $\vec{P} = \kappa \sinh t \,\hat{a}$, and get:
\begin{equation}
\begin{split}
    \frac{1}{ \pi} \int_0^{2 \pi} \sin^{2}{\theta} \rd \theta \int_{\cH^{+}\cup\cH^{-}} \frac{\rd^{2} \hat{u}}{4 \pi}
    & = 
    \frac{1}{\pi^{2} \kappa^{3}} \int_{|P|< \kappa} \frac{\rd^{3} \vec{P}}{\sqrt{1 - \frac{\vec{P}^{2}}{\kappa^{2}}}}
    \\
    \frac{1}{ \pi} \int_0^{\infty} \sinh^{2}t \; \rd t \int_{\cH} \frac{\rd^{2} \hat{a}}{4 \pi}
    & = \frac{1}{\pi^{2} \kappa^{3}} \int_{\mathbb{R}^{3}} \frac{d^{3} \vec{P}}{\sqrt{1 - \frac{\vec{P}^{2}}{\kappa^{2}}}}
    \,.
\end{split}
\end{equation}

%%%%%%%%%
\section{Distributions on SU(2) and SO(3)}
\label{appendixB}
%%%%%%%%%

In this appendix, we describe some properties of the distribution $\delta_\varphi$ and $\Tilde{\delta}_\varphi$, which play the role of Hadamard propagator. We also detailed on-shell one-loop computation. Expression of those distributions is given by
\begin{equation}
    \delta_\varphi(g) = \sum_{j \in \mathbb{N}/2} \chi_j(\varphi) \chi_j(g) = \frac{\pi}{2 \sin\phi \sin\theta}[\delta(\theta - \phi) - \delta(\theta + \phi)]
    \,, \qquad \Tilde{\delta}_\varphi(g) = \frac{\delta_\varphi(g) + \delta_{\pi - \varphi}(g)}{2}
    \,.
\end{equation} \label{equ:delta}
Those distributions verify the identity below: 
\begin{equation}
    \int_{\SU(2)} \rd g \; \delta_\varphi(g) F(g) = \int_{\cS^{2}} \frac{\rd^{2} \hat{u}}{4 \pi} \,F(\varphi, \hat{u})
    \,.
\end{equation}
%This can be easily show, expression $\delta_\varphi(h)$ with dirac distribution \ref{equ:delta} and the explicite expression of the SU(2) Haar measure given in \ref{appendixA}.
Using this distribution, it is straightforward to compute the one-loop on-shell integral for elliptic group elements $g(\theta, \hat{u})$, with angles $\theta \in [- \pi, \pi]$, and $\varphi \in [0, \pi/2]$:
\begin{equation}
    \int_{\SU(1,1)}\rd h \; \delta_\varphi(h) \delta_\varphi(hg)
    = \int_{\cS^{2}} \frac{\rd^{2} \hat{v}}{4 \pi} \delta_\varphi(h(\varphi, \hat{v})g) = \left\{
    \begin{array}{ll}
         \frac{\pi}{4 \sin^{2}(\varphi) \sin(\theta)}  & \quad \text{ if } |\theta| \leq 2 \varphi\\
        0 & \quad \text{ else } 
    \end{array}
\right.
\label{distribSU(2)}
\end{equation}
To prove this, we write the product $hg$ for groups elements $g(\theta, \hat{u})$ and $h(\varphi, \hat{v})$~: 
\begin{equation}
hg
=
\big{[}\cos\theta\cos\varphi - \sin\theta \sin\varphi \hat{u} \cdot\hat{v}\big{]}
\,\id
+ i
\big{[}
\sin\theta \cos\varphi \hat{u} + \cos\theta \sin\varphi \hat{v} + \sin\theta) \sin\varphi \hat{u}\times \hat{v}
\big{]}\cdot\vec{\sigma}
\,.
    \label{EuclidAngleRelation}
\end{equation}
We compute  the rotation angle  $\lambda$ of $hg$  in terms of the angle between the rotation axis of the two group elements, $\cos\beta = \hat{u}\cdot \hat{v}$ :
\begin{equation*}
    \cos\lambda = \cos\theta\cos\varphi - \sin\theta \sin\varphi \cos\beta = \cos(\theta - \alpha) \sin^{2}\frac{\beta}{2} + \cos(\theta + \alpha) \cos^{2}\frac{\beta}{2}
    \,.
\end{equation*}
%This allows to  change variables between $\beta$ and $\lambda$. Using the explicit expression of the $\SU(2)$ Haar measure given in \eqref{measureSU(2)}
This allows to  compute the change of variables between $\beta$ and $\lambda$:
\begin{equation*}
    \frac{1}{\pi} \int_0^{\pi} \sin^{2}\varphi\rd \varphi \int_0^{\pi} \sin\beta \rd \beta \int_0^{2 \pi} \rd \gamma
    = 
    \frac{1}{\pi} \int_0^{\pi} \sin^{2}\varphi \rd \varphi \int_{\varphi - \theta}^{\varphi + \theta} \frac{\sin\lambda}{\sin\theta\sin\varphi} \rd \lambda \int_0^{2 \pi} \rd \gamma
    \,.
\end{equation*}
Then the distribution $\delta_\varphi(hg)$  fixes the angle $\lambda$ of the group element $h(\varphi,\hat{v})g$ to $\varphi$, so we get the condition:
\begin{equation*}
    \hat{u}\cdot\hat{v}  = \frac{\cos{\phi} (\cos{\theta} - 1)}{\sin{\phi} \sin{\theta}} = - \frac{\tan{\frac{\theta}{2}}}{\tan{\phi}}\,.
\end{equation*}
Since $\hat{u}\cdot\hat{v} \leq 1$, this expression is only valid  when $|\theta| \leq 2 \varphi$. Putting all results together, this proves \ref{distribSU(2)} and \eqref{equ:SU(2)onshell}.

Computation of the character convolution works the same way:
\begin{equation*}
    \int_{\SU(2)} \rd h  \; \chi_j(h) \chi_k(hg) =  \frac{1}{\pi} \int_0^{\pi} \rd \varphi \;  \sin^{2}\varphi  \int_0^{\pi} \rd \beta \; \sin\beta\frac{\sin d_j \varphi}{\sin\varphi}\frac{\sin d_k \lambda}{\sin\lambda} = \frac{1}{\pi \sin\theta} \int_0^{\pi} \rd \varphi \; \sin(d_j \varphi)  \int_{\varphi - \theta}^{\varphi + \theta} \rd \lambda \; \sin d_k \lambda
    \,.
\end{equation*}
Those two integrals can now be computed and the result reads:
\begin{equation}
    \int_{\SU(2)} \rd h  \; \chi_j(h) \chi_k(hg) = \frac{\chi_j(g) \delta_{jk}}{d_j}
    \,.
\end{equation}
This expression for characters convolution is useful for tone-loop computations.
%In $\SO(3)$ we find same result but for $j \in \mathbb{N}$ and the integration on $\int_{\SO(3)}$

%%%%%%
\section{Distributions on SU(1,1) and SO(2,1)}
\label{app:distriL}
\label{appendixC}
%%%%%%

In the Lorentzian theory in order to build the Hadamard propagator we first consider the Wightman propagators.We will give their expression for ellitique elements, $g(\theta, \hat{u})$:
\begin{equation}
    \begin{split}
    \delta_\varphi(g)&  = \sum_{j \in  \mathbb{N}/2} [\chi^{+}_j(\varphi) \chi^{-}_j(g) + \chi^{-}_j(\varphi) \chi^{+}_j(g)]
    = \frac{1}{4 \sin\varphi\, \sin\theta} \Big{[}2 \pi \delta(\theta - \varphi) - 1\Big{]} \\
    \Tilde{\delta}_\varphi(g)&   =\sum_{j \in  \mathbb{N}} [\chi^{+}_j(\varphi) \chi^{-}_j(g) + \chi^{-}_j(\varphi) \chi^{+}_j(g) ]= \frac{\pi}{ 4 \sin\theta\, \sin\varphi} \Big{[}\delta(\theta - \varphi) - \delta(\theta - \varphi- \pi)\Big{]}  = \frac{1}{2}\Big{(}\delta_\varphi(g) + \delta_{\varphi + \pi}(g)\Big{)}
    \end{split}    
\end{equation}
In the following part we note  $h(\alpha, \hat{v}) = h(\alpha, t, \gamma),$ with $ \alpha \in [0, \pi], \; t \in [0, \infty[$ and $\gamma \in [0,2 \pi]$ .The positive (resp. negative) Wightman distribution $\delta_\varphi(h)$ (resp. $\delta_{-\varphi}(h)$) has the properties to fix $\hat{v}$ in the upper (resp. lower) hyperboloid and to fix the angle $\alpha = \varphi$ (resp. $\alpha = - \varphi$). we have the following result \cite{Freidel:2005bb}:
\begin{equation}
    \int_{SO(2,1)} dh \; \Tilde{\delta}_\varphi(h) F(h) = \int_{\cH^{+}} dx \; F(x h_\varphi x^{-1}) \quad \int_{SO(2,1)} dh \; \Tilde{\delta}_{-\varphi}(h) F(h) = \int_{\cH^{-}} dx \; F(x h_{-\varphi} x^{-1}) 
\end{equation}
where $h_{\pm \varphi} = \cR(\pm \varphi)$ is an element of the Cartan subgroup of rotation. Explicitly, we have the formula:
\begin{equation}
    \int_{\SO(2,1)} dh \; \Tilde{\delta}_{\pm \varphi}(h) F(h) = \frac{1}{8 \pi}\int_{\cH^{\pm}} d^{2} \hat{v} \; F(h(\pm \varphi, \hat{v}) )
\end{equation} 
\vspace{2.5ex}

%%%
\subsection{Computation of the $\SO(2,1)$ on-shell one-loop integral}
%%%

For the on-shell one loop computation, let us take $g(\theta, \hat{u}) = g(\theta, s, \beta)$ with $\theta \in [0, \pi/2], \; s \in [0, \infty[$, $h(\alpha, \hat{v}) = h(\alpha, t, \gamma),$ with $ \alpha \in [0, \pi], \; t \in [0, \infty[$. We arbitrary chose $g \in \cH^{-}$ but calculations are similar for $g \in \cH^{+}$. In a first time take $\varphi$ such that $2 \varphi \in [0, \pi/2]$. With properties of Wightman distribution, we can simplify the one-loop expression and cut it out into four parts:\begin{equation}
    \begin{split}
        & J_1 = \frac{i \tan^{2}\varphi}{2 \pi \kappa}  \int_{\mathcal{H}^{+}} \frac{d \hat{u}}{8 \pi} [\delta_\varphi(h(\varphi)g)  + \delta_{-\varphi}(h(\varphi)g) \qquad
         J_2  = \frac{i \tan^{2}\varphi}{ 2\pi \kappa}  \int_{\mathcal{H}^{+}} \frac{d \hat{u}}{8 \pi} [\delta_{\pi + \varphi}(h( \varphi )g) + \delta_{\pi -\varphi}(h( \varphi )g)]\\
        & J_3  = \frac{i \tan^{2}\varphi}{2 \pi \kappa}  \int_{\mathcal{H}^{-}} \frac{d \hat{u}}{8 \pi} [\delta_\varphi(h(-\varphi)g)  + \delta_{-\varphi}(h(-\varphi)g) \quad
         J_4  = \frac{i \tan^{2}\varphi}{2 \pi \kappa}  \int_{\mathcal{H}^{-}} \frac{d \hat{u}}{8 \pi} [\delta_{\pi + \varphi}(h( -\varphi )g) + \delta_{\pi -\varphi}(h( -\varphi )g)]\\
    \end{split}
\end{equation}

An interesting property of unitary timelike vectors in Lorentzian space is $|\hat{u} \cdot\hat{v}| \geq1$. If $\hat{u}$ and $\hat{v}$ are in the same hyperboloid, $\hat{u}\cdot \hat{v}\geq 1$, else $\hat{u}\cdot \hat{v} \leq - 1$.
\begin{itemize}
    \item $J_1$ \textbf{computation:} This integral is non-zero if the product $hg$ is elliptic. Thus we can parametrize $(hg)(\lambda, \hat{w})$ with angle $\lambda \in [0, \pi/2]$ and unit vector $\hat{w}^{2} = +1$. Furthermore $\hat{u}$ and $\hat{v}$ are in different hyperboloids and  we can express $\hat{u}$ and $\hat{v}$ with the angle $\lambda, \theta, \alpha$, thus:
\begin{equation}
    \hat{u}\cdot \hat{v} = \frac{\cos\theta \cos\alpha + \cos\lambda }{\sin\theta \sin\alpha} \quad \text{and} \quad \hat{u} \cdot \hat{v} < -1
\end{equation}
 For $J_1$, $\alpha = \varphi$ and $\lambda$ is fix with the dirac distribution to $\varphi$ or $-\varphi$, therefore we get:
 \begin{equation*}
     \hat{u}\cdot \hat{v} = - \f{\tan\f\theta2} {\tan\varphi} \leq -1 \quad  \Longrightarrow \quad \theta \geq 2 \varphi
     \,.
 \end{equation*}
This condition on the angle $\theta$ and $\varphi$ must be fulfilled to have $J_1$ no zero. Keeping this in mind, we now compute explicitely the integral:
\begin{equation*}
    J_1  = \frac{i \tan^{2}\varphi}{8 \kappa \pi} \int_0^{\infty}\rd t \; \sinh t  \frac{2\pi \Big{[}\delta(\lambda - \varphi) - \delta(\lambda + \varphi)\Big{]}}{4 \sin\varphi \sin\lambda}
\end{equation*}
Explicit computation of the integral require a variable change between $\sinh t$ and $\sin \lambda$. First, we need to know more about the hg product of hyperbolic elements. Let us recall that we parametrize this elements by, $g(\theta, \hat{u}) = g(\theta, s, \beta)$, $h(\varphi,\hat{v}) = h(\varphi, t, \gamma)$ and $(hg)(\lambda, \hat{w})$. Relation between those elements reads: 
\begin{equation}
    \cos \lambda = \cos \theta \cos \varphi - \sin \theta \sin \varphi \; \hat{u}\cdot\hat{v}
    \,.
\end{equation}
Here $\hat{u}$ and  $\hat{v}$ belong to two different hyperbola, thus $\hat{u} \cdot\hat{v} \leq - 1$ and we can take parametrize them with $t \in [0, \infty] $ and get $\hat{u} \cdot\hat{v}=-\cosh t$:
\begin{equation*}
    \cos{\lambda} = \cos{\theta} \cos{\varphi} - \sin{\varphi} \sin{\theta} (- \cosh t)
    \,.
\end{equation*}
In the limit $t = 0$, we have $\cos\lambda = \cos(\theta - \varphi)$. On the other hand,  for large $t$, we get $\cos\lambda \to 1$. %(because $\theta, \varphi \in [0,\pi/2])$
 Doing the variable change between $t$ and $\lambda$, one gets :
\be
J_1 = \frac{i}{16 \kappa \sin\theta \cos^{2}\varphi} \int_{\varphi - \theta}^{0}(- d \lambda) [\delta(\lambda - \varphi) - \delta(\lambda + \varphi)]
%\quad \text{&} \quad \theta \geq 2 \varphi
=
\left\{
    \begin{array}{ll}
        0 & \quad \text{if} \quad  \theta \in [0, 2 \varphi]
\vspace*{1mm}        \\
        \displaystyle\frac{i}{16 \kappa \sin\theta\cos^{2}\varphi } & \quad \text{if} \quad  \theta \in [2 \varphi, \pi/2]
    \end{array}
\right.
\ee
    \item $J_2$ \textbf{computation}:  Now $\lambda $ is fixed to $\pi \pm \varphi$ and the scalar product between $\hat{u}$ and $\hat{v}$ read:
    \begin{equation}
        |\hat{u} \cdot\hat{v}| = \frac{1}{\tan\f\theta2 \tan\varphi}
    \end{equation}
     However for $\theta/2, \varphi \in [0,\pi/4]$ this scalar product between $\hat{u}$ and $\hat{v}$ is positive, while $\hat{u}\hat{v}$ should be negative for $h$, $g$ in different hyperboloid. This means the angle for the product $hg$ does not exist, and $J_2$ is null.
    \item \textbf{$J_3$ and $J_4$ computation}: Those two integrals can be computed in the same way than the other one, with $J_3 =0$ and $ J_4 = i /(16 \kappa^{2} \sin\theta \cos^{2}\varphi)$ if $\theta \in [2 \varphi, \pi/2]$.
\end{itemize}

The final result reads, for $\theta \in [0,\pi/2]$ and $\varphi \in [0,\pi/4]$:
\begin{equation}
    \cL_{\varphi}^{0, \rL} = \left\{
    \begin{array}{ll}
        0 & \quad \text{if} \quad  \theta \in [0, 2 \varphi]\\
        \frac{i}{8 \kappa \cos^{2}\varphi \sin\theta} & \quad \text{if} \quad  \theta \in [2 \varphi, \pi/2]
    \end{array}
\right.
\end{equation}
When the mass $\varphi$ increase and lie inside the interval $[\pi/4, \pi/2]$, we have still $J_2 = 0$, but the integral $J_1$ become also zero, because the condition $\theta \geq 2 \varphi$ is no more longer valid for $\theta \in [0, \pi /2]$. Therefore:
\begin{equation}
    \cL_\varphi^{0,\rL}(g) = 0 \quad \forall \varphi \in [\pi/4, \pi/2]
\end{equation}
Those results on $\SO(2,1)$ agree with the one found in \cite{Sasai:2009jm}.

%%%
\subsection{Computation of the $\SU(1,1)$ on-shell one-loop integral}
%\textbf{SU(1,1) on-shell one-loop}
%\vspace{2.5ex}
%%%

Finally we compute the SU(1,1) on-shell one-loop. Let us take $g(\theta, \hat{u}) = g(\theta, s, \beta) \in \cH^{-}, \theta \in [0,\pi], s \in [0,\infty[, \beta \in [0, 2 \pi]$, we parametrize the integration elements $h(\alpha, \hat{v}) = h(\alpha, t, \gamma), \alpha \in [0, \pi], t \in [0, \infty[, \gamma \in [0,2\pi]$. We note $\lambda_+$ the angle of the product $hg$, with $[h( + \varphi)g(\theta)](\lambda_+)$ and  $[h(-\varphi)g(\theta)](\lambda_-)$. We set $\varphi \in [0, \pi]$. With those notations, the SU(1,1) on-shell one-loop reads:
\begin{equation*}
    \begin{split}
        \ell_\varphi^{\rL, 0} & = \frac{i \sin^{2}(\varphi)}{ \pi \kappa}  \int dh \; [\delta_\varphi(h) + \delta_{-\varphi}(h)][\delta_\varphi(hg) + \delta_{-\varphi}(hg) ] \\
        & = \frac{i \sin^{2}(\varphi)}{ 2\kappa \pi}  [\int_{\cH^{+}} \frac{d^{2} \hat{u}}{4 \pi}  + \int_{\cH^{-}} \frac{d^{2} \hat{u}}{4 \pi} ] [ \delta_\varphi(h(\varphi, \hat{u})g) + \delta_{-\varphi}(h(\varphi, \hat{u})g)- \delta_\varphi(h(-\varphi, \hat{u})g) - \delta_{-\varphi}(h(-\varphi, \hat{u})g)] \\
        & = \frac{i \sin(\varphi)}{4 \kappa }[\int_{\cH^{+}} \frac{d^{2} \hat{u}}{4 \pi}  + \int_{\cH^{-}} \frac{d^{2} \hat{u}}{4 \pi} ] ( \delta(\lambda_+ - \varphi) - \delta(\lambda_+ + \varphi) + \delta(\lambda_{-} - \varphi) - \delta(\lambda_{-} + \varphi) 
    \end{split}
\end{equation*}
As for the SO(2,1) on-shell one-loop we will do variable change in this expression between $t$ and $\lambda_\pm$.

\begin{itemize}
    \item For $\hat{u} \in \mathcal{H}_{-}$ relation between angles are given by:
\begin{equation}
    \cos(\lambda_\pm) = \cos(\theta) \cos(\pm \varphi) - \sin(\theta) \sin(\pm \varphi) (- \cosh(s))
    \label{LorentzAngleRela-}
\end{equation}
For $\lambda_+$, one gets $- \sin(\lambda_+) d \lambda_+ =\sin(\theta) \sin(\varphi)  \sinh(s) ds$ and $\lambda_+ \in [0, \varphi-\theta]$

\noindent For $\lambda_-$, one gets $- \sin(\lambda_-) d \lambda_- = - \sin(\theta) \sin( \varphi)  \sinh(s) ds$ and $\lambda_- \in [\pi, \varphi+\theta]$
%%%%%%%%%%%%%%%%%%%%%%%%%%%%%%%%%%%%%%%-
    \item For $\hat{u} \in \mathcal{H}_{+}$ relation between angle are given by:
\begin{equation}
    \cos(\lambda_\pm) = \cos(\theta) \cos(\pm \varphi) - \sin(\theta) \sin(\pm \varphi) \cosh(s)
    \label{LorentzAngleRela+}
\end{equation}
For $\lambda_+$, one gets $- \sin(\lambda_+) d \lambda_+ = - \sin(\theta) \sin(\varphi)  \sinh(s) ds$ and $\lambda_+ \in [\pi, \varphi+\theta]$

\noindent For $\lambda_-$, one gets $- \sin(\lambda_-) d \lambda_- =  \sin(\theta) \sin( \varphi)  \sinh(s) ds$ and $\lambda_- \in [0, \varphi-\theta]$
\end{itemize}
Doing variable change gives us the following expression for the off-shell one loop:
\begin{equation}
\ell_\varphi^{\rL, 0}
=
\frac{i }{4 \kappa \sin\theta }
\int_{[0,\varphi- \theta]\cup[\pi,\varphi + \theta]} \rd\lambda\,
\Big{[} \delta(\lambda - \varphi) -  \delta(\lambda + \varphi)\Big{]}
=
\left\{ \begin{array}{ll}
       0 & \quad \text{if} \quad \theta \in [0, 2 \varphi]
       \,,\\
\displaystyle        \frac{i}{4 \kappa \sin\theta} & \quad \text{if} \quad \theta \in [ 2 \varphi, \pi]\,.
    \end{array}
\right.
\end{equation}
%Finally, after computations, we obtain
%\begin{equation}
%     \ell_\varphi^{\rL, 0}=  \left\{ \begin{array}{ll}
%       0 & \quad \text{if} \quad \theta \in [0, 2 \varphi]\\
%\displaystyle        \frac{i}{4 \kappa \sin\theta} & \quad \text{if} \quad \theta \in [ 2 \varphi, \pi]
%    \end{array}
%\right.
%\end{equation}
A similar result can be derived with $g \in \mathcal{H}^{+}$, leading the expected result \eqref{equ:SU(1,1)onshell} for the on-shell one-loop integral on $\SU(1,1)$.

\bibliographystyle{bib-style}
\bibliography{biblio}

\end{document}